%% file: PTAbs.tex
\documentclass[preprint,amssymb,superscriptaddress]{revtex4-1}

\usepackage{booktabs}
\usepackage[columnwise]{lineno}
\usepackage[usenames,dvipsnames]{xcolor}

\usepackage[caption=false]{subfig}
\usepackage[utf8]{inputenc}
\usepackage{graphicx}
\usepackage{mathtools}
\usepackage{cases}

\usepackage{ulem}
\usepackage{soul}
\usepackage{siunitx}
\usepackage{notes2bib}
\DeclareSIUnit[number-unit-product = {}]\diopter{D}

\input{values}

\newif\iftesting
\newif\ifshortversion
\iftesting \linenumbers \fi                                                 

\newcommand{\testingAbstract}[1]{\iftesting {\bf CURRENTLY IN DRAFT MODE: includes line numbers, TOC, page breaks between sections.} \else #1\fi}         

\newcommand{\modesSym}{E}

\begin{document}
\title{Point absorbers in  Advanced LIGO}

\iftesting \author{Aidan F. Brooks et al} \else \input{authorList_PRD} \fi








\begin{abstract}
Small, highly absorbing points are randomly present on the surfaces of the main interferometer optics in Advanced LIGO. 
The resulting nano-meter scale thermo-elastic deformations and substrate lenses from these micron-scale absorbers significantly reduces the sensitivity of the interferometer  directly though a reduction in the power-recycling gain  and indirect interactions with the feedback control system.
We review the expected surface deformation  from point absorbers and  provide a pedagogical description of the impact on power build-up in second generation gravitational wave detectors (dual-recycled Fabry-Perot Michelson interferometers).
This analysis predicts that the power-dependent reduction in interferometer performance will significantly degrade maximum stored power by up to 50\% and hence, limit GW sensitivity, but suggests system wide corrections that can be implemented in current and future GW detectors. This is particularly pressing given that future GW detectors call for an order of magnitude more stored power than  currently used in Advanced LIGO in Observing Run 3. We briefly review strategies to mitigate the effects of point absorbers in current and future GW wave detectors  to maximize the success of these enterprises. 
\testingAbstract{}



\end{abstract}

\maketitle

\section{Introduction}
\input{intro}

\section{Modeling formalism}
\input{modeling_summary}

\section{Observation of point absorber effects in Advanced LIGO}
\label{sec:observations}
\subsection{Optical gain vs power and position}
\input{reductionInPRGwPower.tex}

\subsection{Time evolution of PRG reduction}

\input{TimeConstantOfPRG.tex}

\subsection{Noise couplings from ITM thermal lens effects}
\input{ITMnoise.tex}


\section{Considerations for working with point absorbers in the future}
\input{FutureIFOs}
\section{Conclusion}
\input{conclusion.tex}

\section{Acknowledgements}
The authors gratefully acknowledge the support of the United States
National Science Foundation (NSF) for the construction and operation of the
LIGO Laboratory and Advanced LIGO as well as the Science and Technology Facilities Council (STFC) of the
United Kingdom, and the Max-Planck-Society (MPS) for support of the construction of Advanced LIGO.
Additional support for Advanced LIGO was provided by the Australian Research Council.
The authors acknowledge the LIGO Scientific Collaboration Fellows program for additional support.

LIGO was constructed by the California Institute of Technology and Massachusetts Institute of Technology with funding from the National Science Foundation, and operates under cooperative agreement PHY-1764464. Advanced LIGO was built under award PHY-0823459. This paper carries LIGO Document Number LIGO-P1900287.

\appendix

\ifshortversion  \else
\section{Advanced LIGO interferometer operation}
\input{aLIGOOperation}
\fi



\section{Thermal-elastic surface deformation from point absorbers}
\input{optics_and_pt}
\section{Single-bounce deformation-induced scattering into higher order modes}
\input{single_bounce_scattering}
\section{Resonant optical interaction, excess losses and optical gain}
\input{IFOproperties}

\input{HOM_resonance_discussion}

\input{PRG_discussion}

\ifshortversion  \else 
\section{Substrate thermal lens, noise couplings and impacts on control systems}
\input{ITMeffects}
\fi

\section{Table of parameters}
\label{sec:table}

\input{parameters}

\bibliographystyle{TCS}
\bibliography{PTAbs}

\end{document}

%% file: values.tex
\providecommand\DefDataVal[2]{%
  \expandafter\newcommand\csname DataVal-#1\endcsname{#2}%
}
\providecommand{\DataVal}[1]{\csname DataVal-#1\endcsname}
\DefDataVal{AverageSpotAbsorption}{$100-350$\,ppb} 
\DefDataVal{UniformAbsorption}{500$\pm$250\,ppb}  
\DefDataVal{ArmCavityFinesse}{280}

\DefDataVal{CCpole}{\SI{0.6}{\hertz}}
\DefDataVal{ASPortTransmission}{10-20ppm}

\DefDataVal{MaxArmPower}{\SI{750}{\kilo\watt}}
\DefDataVal{MaxInputPower}{\SI{125}{\watt}}

\DefDataVal{ITMTrans}{1.4\%}

\DefDataVal{TEM00LineWidth}{0.24\%}
\DefDataVal{TEM60LineWidth}{1.75\%}
\DefDataVal{TEM70LineWidth}{6.45\%}

\DefDataVal{H1ITMXPabs}{$27\pm2.5$\,mW}  
\DefDataVal{MM2ArmLossRatio}{3 to 11 times} 
\DefDataVal{MM2ArmLossRatioWFinesse}{28 to 95 times} 

%% file: authorList_PRD.tex
\author{A.~F.~Brooks}
\affiliation{LIGO, California Institute of Technology, Pasadena, CA 91125, USA}
\author{G.~Vajente}
\affiliation{LIGO, California Institute of Technology, Pasadena, CA 91125, USA}
\author{H.~Yamamoto}
\affiliation{LIGO, California Institute of Technology, Pasadena, CA 91125, USA}
\author{R.~Abbott}
\affiliation{LIGO, California Institute of Technology, Pasadena, CA 91125, USA}
\author{C.~Adams}
\affiliation{LIGO Livingston Observatory, Livingston, LA 70754, USA}
\author{R.~X.~Adhikari}
\affiliation{LIGO, California Institute of Technology, Pasadena, CA 91125, USA}
\author{A.~Ananyeva}
\affiliation{LIGO, California Institute of Technology, Pasadena, CA 91125, USA}
\author{S.~Appert}
\affiliation{LIGO, California Institute of Technology, Pasadena, CA 91125, USA}
\author{K.~Arai}
\affiliation{LIGO, California Institute of Technology, Pasadena, CA 91125, USA}
\author{J.~S.~Areeda}
\affiliation{California State University Fullerton, Fullerton, CA 92831, USA}
% ...

\collaboration{The LIGO Scientific Collaboration}



%% file: intro.tex

\begin{figure}
    \centering
  \includegraphics[width=\columnwidth]{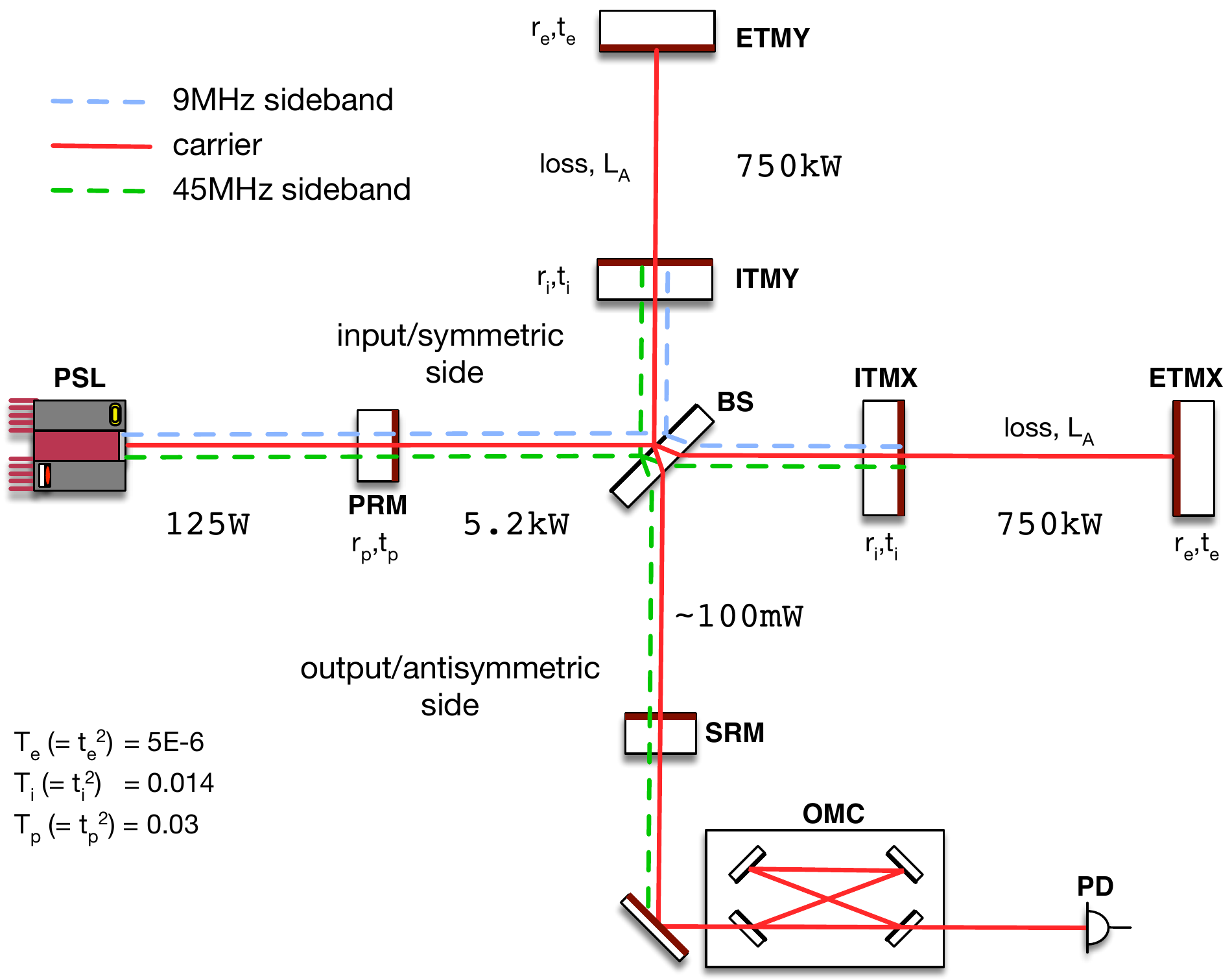}
  \caption{Schematic diagram of Advanced LIGO showing the main optics, nominal power levels and different optical frequencies present in the interferometer. 
  The detector is a Fabry-Perot Michelson, with power-recycling and signal-recycling cavities. The power recycling cavity (PRC) and Fabry-Perot arm cavities hold large amounts of stored power. At \SI{125}{\watt} of input laser power, the nominal design yields \SI{5.2}{\kilo\watt} stored in the PRC and \SI{750}{\kilo\watt} in the arms. The output mode cleaner (OMC) is designed to transmit only the fundamental spatial mode of the arms, which carries the GW signal, and reject other spatial modes. Radio-frequency sidebands are injected into the detector to measure and control the multiple degrees of freedom created by the coupled resonant cavities.}
  \label{fig:IFO_schematic}
\end{figure}

The Advanced LIGO (aLIGO) gravitational wave (GW) detectors, in conjunction with the Virgo GW detector,  having  completed their third observing run (O3)\cite{O3CommissioningPRD},  have reported detection of multiple gravitational wave events \cite{GW150914etal, O1O2Catalog} (10 binary black-hole mergers and 1 binary neutron star mergers) and issued 56 public alerts for detection candidates \footnote{https://gracedb.ligo.org/superevents/public/O3/}. 
The aLIGO detectors, illustrated in Figure \ref{fig:IFO_schematic}, are high laser power, dual-recycled, \SI{4}{\kilo\meter} Fabry-Perot, Michelson interferometers operating at \SI{1064}{\nano\meter} \cite{aLIGO_Overview}. 
Passing gravitational waves cause a strain in space-time, resulting in a differential length variation of the Fabry-Perot arms that yields a detectable  intensity variation at the output port of the interferometer. 

The sensitivity of the interferometers is limited by a variety of technical and fundamental sources.  At  frequencies above approximately \SI{100}{\hertz} quantum noise, in the form of shot noise on the photodetector, is the limiting noise source \cite{Corbitt_2004}.
Quantum noise can be decreased by increasing the input laser power injected into the interferometer, thereby increasing the amount of stored power in the interferometer arms. The design level for aLIGO is \DataVal{MaxInputPower}{} input power with a stored arm power of  \DataVal{MaxArmPower}{} per arm as illustrated in Figure \ref{fig:IFO_schematic}. During O3, Advanced LIGO routinely operated at input power levels of \SI{35}{\watt} $\--$ \SI{40}{\watt} \cite{O3CommissioningPRD}.

Optical power is absorbed at the sub-ppm level in the surfaces of the main LIGO optics, referred to as the Test Masses (ITMX, ITMY, ETMX and ETMY in Figure \ref{fig:IFO_schematic}).  
As stored power is increased, these optics are exposed to several hundred kW of resonating power, they absorb several tens of mW causing thermo-elastic deformation of the optical surfaces and thermo-refractive lenses in the substrates \cite{PhysRevA.44.7022, HelloVinet_TE:90, HelloVinet:90_TL}. 

Absorption is classified as uniform and non-uniform. {\it Uniform absorption} is characterized by a spatially invariant (or nearly invariant) absorption coefficient across the high-reflectivity (HR) surface of the optic. 
In the Advanced LIGO test masses, the unambiguously measured uniform absorption values ranged over  \DataVal{UniformAbsorption}{}. 
In the case of uniform absorption, non-uniform, low-spatial frequency thermal lenses \cite{HelloVinet:90_TL} and surface deformations \cite{HelloVinet_TE:90} are well approximated by spherical wavefront errors. Advanced LIGO contains a thermal compensation system (TCS)  \cite{Brooks:16} - not shown in Figure \ref{fig:IFO_schematic} - with actuators  to provide low-spatial correction for (a) thermal lenses in the substrates and (b) curvature errors on the surfaces of the test masses.

\begin{figure}[ht]

    \centering
  \includegraphics[width=\columnwidth]{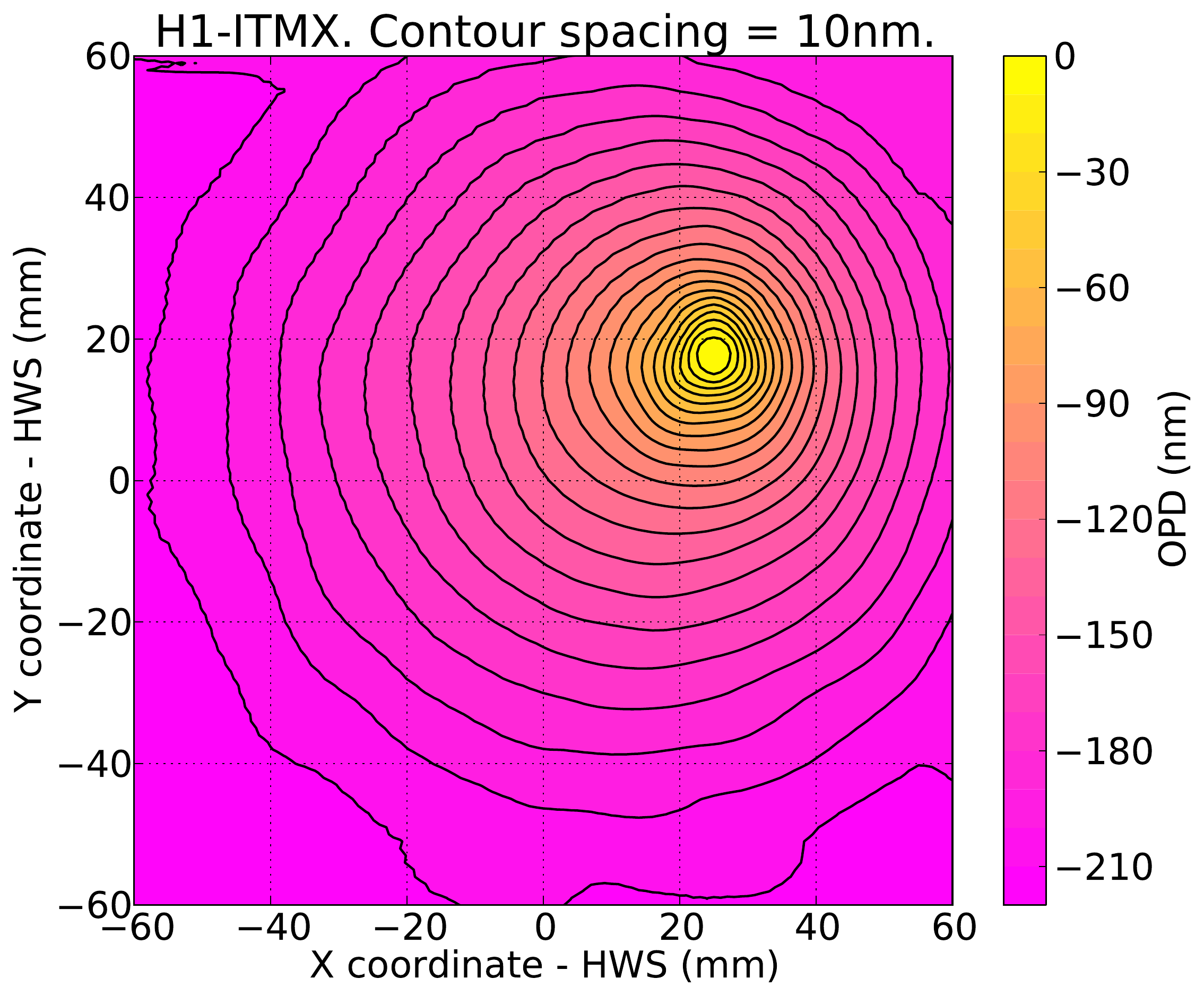}
  \caption{Hartmann sensor measurement of optical path distortion (thermo-refractive plus thermo-elastic) from a single point absorber on H1-ITMX. Cold reference  taken at GPS time: \SI{1180229513}{\second}, hot measurement taken  \SI{3322}{\second} later at GPS time:  \SI{1180232835}{\second}. This measurement corresponds to approximately  \DataVal{H1ITMXPabs}{}  power absorbed in the point.}
  \label{fig:HWS_measurementX}
\end{figure}




{\it Non-uniform absorption} is any form of absorption with high-spatial frequency dependence (where ''high spatial frequency'' refers to  features that are significantly smaller than the Gaussian beam diameter of the illuminating laser beam).  A salient example is a point-like absorber (a ''point absorber''): a sub-millimeter scale, highly absorbing region on the surface of the test mass. 

\begin{figure}
    \centering
  \includegraphics[width=\columnwidth]{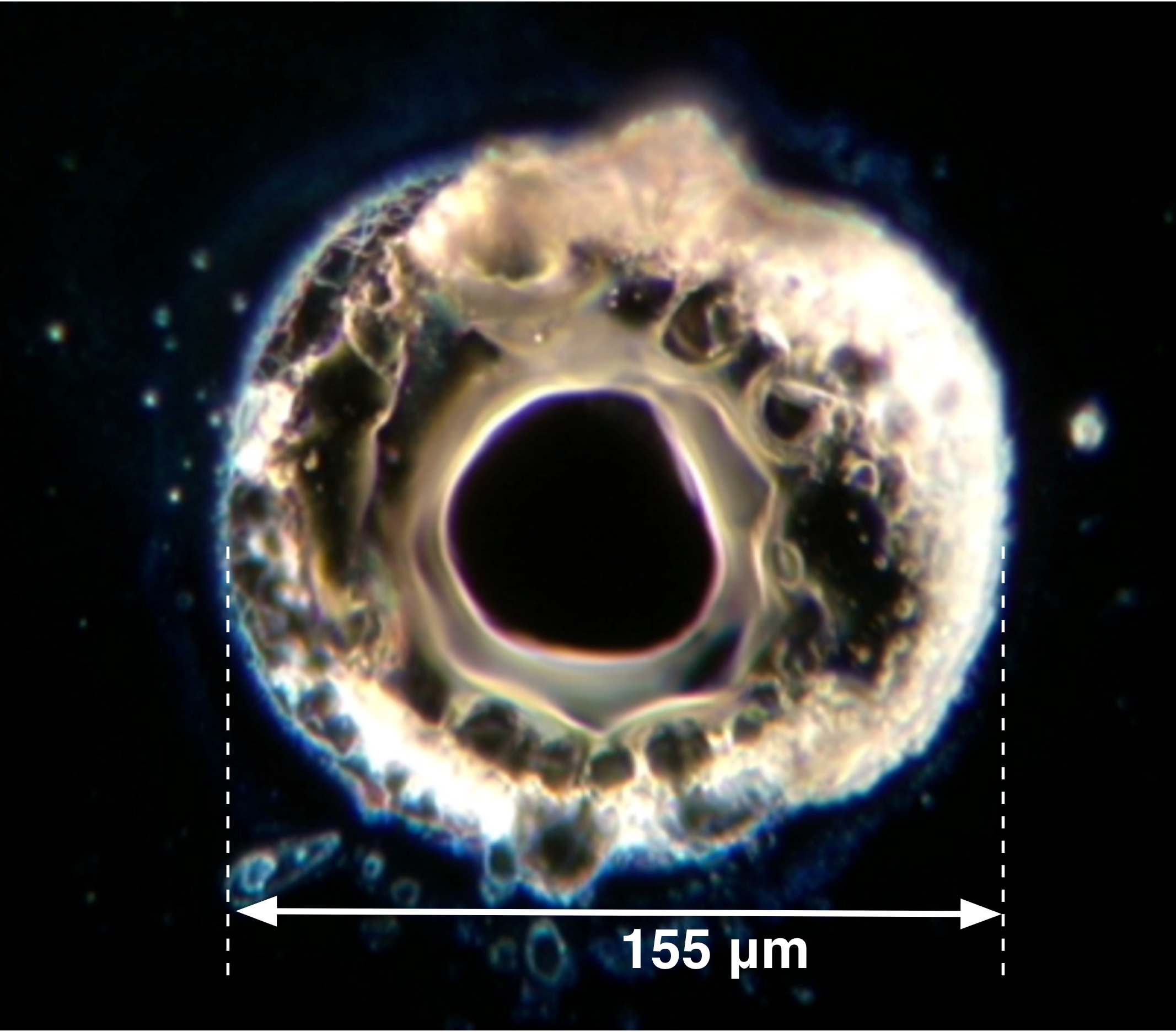}
  \caption{Dark field microscope image of point absorber measured on an Advanced LIGO optic (corresponding to the thermal lens measurement shown in Figure \ref{fig:HWS_measurementX}). Also shown in Buikema et. al. \cite{O3CommissioningPRD}.}
  \label{fig:microscope_pt}
\end{figure}

Within aLIGO, Hartmann  wavefront sensors (HWS)   \cite{Brooks:16, Brooks:09} measure the  spatial distribution of the integrated thermo-refractive and thermo-elastic deformations in the main LIGO optics induced by operation at high power. Measurements  performed in-situ have detected unambiguous evidence of  point absorbers on at least 5 of 8 observed test masses \footnote{{https://dcc.ligo.org/LIGO-G1900693}}. An example of these measurements  is shown in Figure \ref{fig:HWS_measurementX}.
Additional forensics of the aLIGO optics (performed off-site on uninstalled optics)  revealed point absorbers on multiple optics, including several optics never exposed to high laser power in the vacuum system. A microscope image of a point absorber is shown in Figure \ref{fig:microscope_pt}.

The spatial resolution of the Hartmann wavefront sensor measurements is approximately one sample every \SI{7.5}{\milli\meter} in both transverse dimensions. Hence we cannot resolve the features in the wavefront smaller than this. However, due to thermal diffusion to scales larger than the spatial resolution, we can infer  the total  power absorbed  by fitting thermal lens models to this data. 
Observed absorption  values lie within the range \DataVal{AverageSpotAbsorption}{} of the total incident power, however, the number of measurements is too small to reliably describe the distribution of absorption values.
The  thermal lens and surface deformations induced by non-uniform absorption and point absorbers are characterized by features smaller than the incident laser beam size (by high spatial frequencies) and, currently, aLIGO contains no high spatial frequency thermal compensation system to correct for these effects.

The origin of the point absorbers in Advanced LIGO is currently under intense investigation. The absorbers cannot be removed with standard cleaning techniques and appear to be embedded in the coating. Initial spectroscopic analysis of absorbing points on witness samples (coated in same coating runs as aLIGO optics) show high concentrations of aluminum. Additionally, following a recent inspection of an aLIGO optic, more point absorbers appeared on that optic - apparently introduced during the inspection process. The nature of these new contaminants is not clear. The full scope of point absorber forensics is beyond the scope of this paper and will be addressed in a future manuscript. Additionally, it is not clear if the presence of point absorbers is unique to the large scale optics used in gravitational wave detectors or is common in all dielectric layer precision optics. 


As Vajente \cite{Vajente:14} described, non-uniform surface deformation of the optics in a gravitational wave detector will scatter power into higher-order spatial modes (HOM) in the interferometer. The Fabry-Perot arms cavities will enhance and suppress different HOM and may, depending on cavity geometry and other factors discussed here, resonantly extract power from the fundamental mode, increasing the observed loss. We briefly review this in Section \ref{sec:short_modeling}. 


Within this article, we explore the thermo-optical interaction of point absorbers with a high-power, high-finesse, Michelson interferometer. 
Observations of the interferometer response  are reported in Section \ref{sec:observations} and projections for future impacts on Advanced LIGO are presented in Section \ref{sec:future}. 

To provide more context, extended background material is included in the Appendices. The basic operation of aLIGO is described in Appendix \ref{sec:aLIGOOperation}.
The effects of point absorbers on (a) just the thermal state of the optic are described in Appendix \ref{sec:Optics}, (b) the simple laser/optic interaction in Appendix  \ref{sec:single-bounce} and (c) the resonant-cavity/laser/optic interaction in Appendix  \ref{sec:IFOproperties}. 
The impact of substrate thermal lensing on the interferometer control systems (and subsequent noise couplings) is examined in Section \ref{sec:ITMeffects}.

%% file: modeling_summary.tex
\label{sec:short_modeling}
In a dual-recycled, Fabry-Perot Michelson interferometer, the power-recycling gain (the ratio of stored laser power in the power-recycling cavity to input laser power) is a good proxy for monitoring the average loss in the arms of the interferometer.
In the appendices of this manuscript, we describe a formalism for modeling the effective change in power-recycling gain due to point absorbers on the test masses. We summarize the main findings here and interested readers are encouraged to review the appendices for further detail.

As shown in Equation \ref{eqn:outside_abs}, the surface deformation from a point absorber  is approximated by:

\begin{equation}
    \Delta s \approx \frac{\alpha\,P_{\mathrm{abs}} }{2\,\pi\,\kappa} \, f(r)
\end{equation}

\noindent where $f(r)$, the functional form of the radial spatial distortion from a point absorber, is defined in Equation \ref{eqn:fr_defn}. The surface deformation is proportional to the coefficient of thermal expansion, $\alpha$, the absorbed power, $P_\mathrm{abs}$ in a point and inversely proportional to the thermal conductivity, $\kappa$. 
When the fundamental mode in a Fabry-Perot cavity reflects off a mirror distorted by the above surface deformation, it  scatters some of that field into higher order modes (HOM). The cavity  resonantly enhances/suppresses those modes as a function of the round-trip phase they accumulate in the cavity. The loss of power from the fundamental mode to a HOM,  $\mathcal{L}_{mn}$, is approximated by Equation \ref{eqn:Gab_A}, which is reproduced here:

\begin{equation}
    \mathcal{L}_{mn} = a_{00|mn}^2\, g_{mn}. \label{eqn:amp_gain}
\end{equation}

\noindent We emphasize here that losses are a function of two main elements. The first element, $a_{00|mn}$, defined in Equation \ref{eqn:scatt_posn_dependence}, is {\it single-bounce amplitude scattering} from the fundamental mode into the $mn-\mathrm{th}$ HOM 
when reflected off the mirror with the surface deformation $\Delta s$ located at position $\mathbf{r_c}$. The second element is the {\it resonant enhancement/suppression factor} of the HOM for the Fabry-Perot cavity geometry and accounting for additional phase delays experienced by HOM due to surface polish errors at the edges of the mirrors, defined in Equation \ref{eqn:Gab_B}.

The average arm loss experienced by the fundamental mode is the sum of losses across all modes other than the fundamental mode:

 \begin{eqnarray}
      \mathcal{L}_\mathrm{A} & = & \mathcal{L}_\mathrm{nom} + \frac{1}{2} \sum_{m,n}  \mathcal{L}_{mn,X}  + \mathcal{L}_{mn,Y} \label{eqn:loss_avg1}\\
       & = & \mathcal{L}_\mathrm{nom}+ b\, \left(\frac{P_A}{100\mathrm{kW}}\right)^2 \label{eqn:loss_avg2}
       \end{eqnarray}
From this model, the power-recycling gain of the interferometer is   summarized in Equation \ref{eqn:PRG_Psq}, reproduced here:

\begin{equation}
       G_P =  \left( \frac{t_p}{1-r_p\,  \left(1 - \frac{G_{A}}{2}\,\left[\mathcal{L}_\mathrm{nom} + b\, \left(\frac{P_A}{100\mathrm{kW}}\right)^2\right]\right)} \right)^2 \label{eqn:PRGfull}
\end{equation}

In Appendix \ref{sec:prg_generalized_discussion}, we highlight all of the physical elements that contribute to the total loss, either through single-bounce amplitude scattering, $a_{00|mn}$, (specifically total absorbed power, material properties and location of position absorber) or through the resonant enhancement/suppression factor, $g_{mn}$, (namely clipping losses, cavity geometry and mirror polish errors).

As shown in Appendix \ref{sec:single-bounce_size}, the loss into HOM is  dependent only on the absorbed power, and not the size or distribution of the point absorber. Additionally, it is shown in Appendix \ref{sec:HOM_scattering_tau} that time-scale for scattering into higher order modes decreases as mode-order increases - allowing us to use the time-evolution of power-recycling gain as a proxy for the contribution of HOM to loss in the arm.

%% file: reductionInPRGwPower.tex
 
 \begin{figure}
  \centering
  \includegraphics[width=\columnwidth]{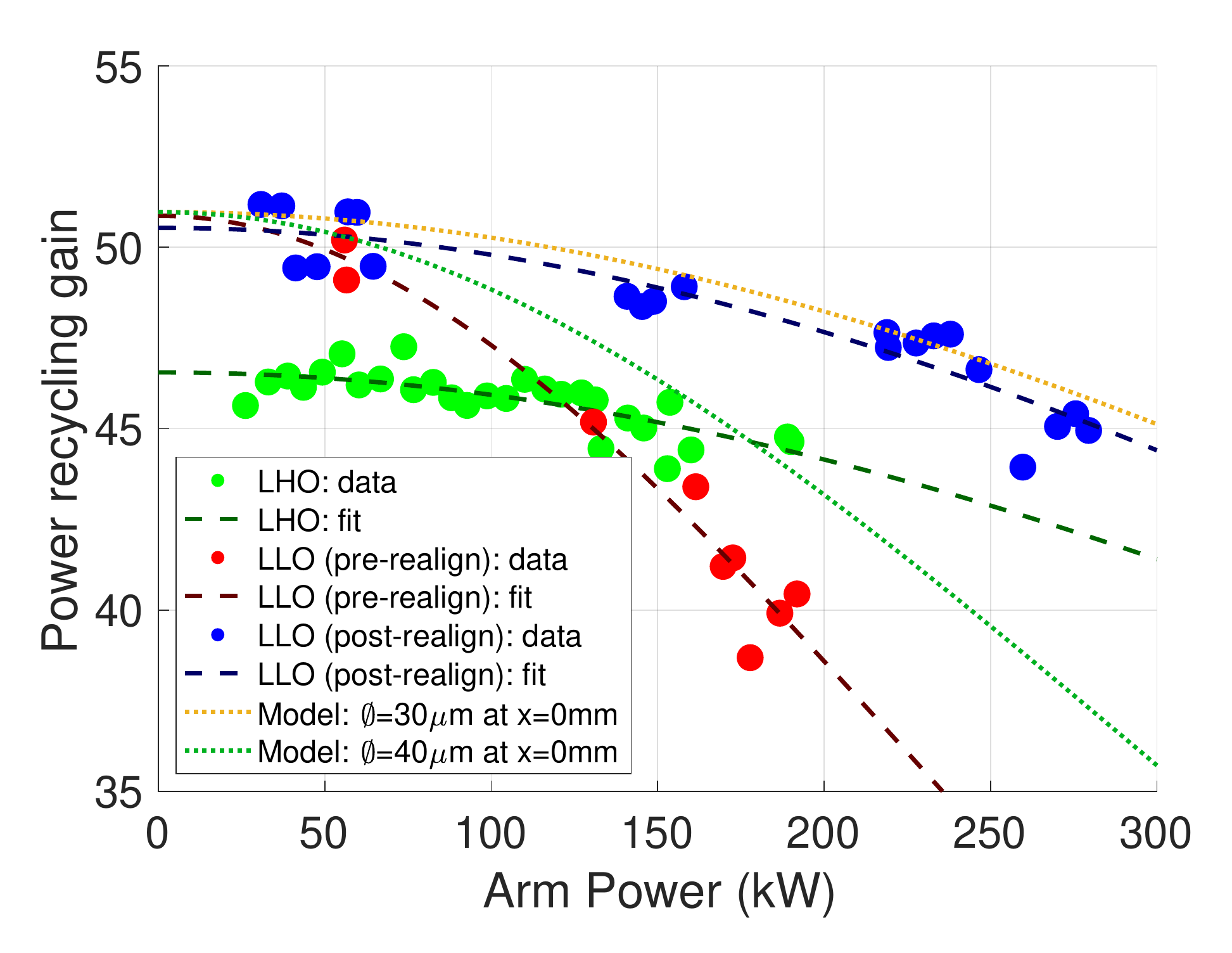}
  \caption{Average LIGO power recycling gain versus arm power for the first 120 days of 2019. The Livingston data has been split into two epochs: pre- and post-interferometer realignment. Arm power is the product of the measured input laser power, the measured power recycling gain, the arm cavity optical gain and the beam-splitter transmission. Models of the PRG are shown by dotted yellow and green curves (assuming  a $\SI{30}{\micro\meter}$ diameter point absorber at $x=0$mm and a $\SI{40}{\micro\meter}$ diameter point absorber at $x=0$mm, respectively). The large range covered by the models indicates a strong variation in the PRG behavior is expected depending on the exact configuration of point absorbers. }
  \label{fig:PRG_v_power}
\end{figure}


In aLIGO, we observed the power recycling gain  decay as a function of input laser power, $P_{\mathrm{in}}$ and arm power. The latter is determined from the product of the power-recycling gain, the input laser power, the arm cavity optical gain and the beam-splitter transmission:

\begin{equation}
    P_A = 0.5\, G_P\, G_A\, P_\mathrm{in}
\end{equation}

For O3, the average arm cavity optical gains, $G_A$, are 268 and 265 for Hanford (LHO) and  Livingston (LLO), respectively (for full details, see Section VB in Buikema et al.\cite{O3CommissioningPRD}). Figure \ref{fig:PRG_v_power}  shows the  power-recycling gain (averaged into \SI{1}{\watt} bins) versus the input laser power over a 120-day period in early 2019 for the two LIGO interferometers, LLO and LHO, with two epochs plotted for the former. 
After an initial commissioning period the Livingston interferometer was realigned to move the arm mode by approximately \SI{30}{\milli\meter} on ETMY \footnote{D. Martynov, {LLO aLOG 43121, February 2019}, {https://alog.ligo-la.caltech.edu/aLOG/ }}. The Livingston data reflects this and is split into two epochs: pre- and post-interferometer realignment,  red and blue data, respectively.  The optimum Hanford alignment is also shown by green data.

Using the  data sets of the measured power recycling gain vs arm power, $P_A$, we fit for $\mathcal{L}_\mathrm{nom}$ and $b$ in Equation \ref{eqn:PRG_Psq}, yielding the following results:

\input{fitTable}


 The new alignment, blue data in Figure \ref{fig:PRG_v_power}, shows a $5\times$  reduction of   power-dependent loss, strongly illustrating the position-dependent arm loss described in Section \ref{sec:single-bounce}.

Lastly, we have modeled maxima and minima for expected power recycling gain, dotted green and yellow, respectively, calculated using the model represented by Equations \ref{eqn:loss_avg1}, \ref{eqn:loss_avg2} and \ref{eqn:PRGfull}, assuming absorption ranges of 100-350ppb, varying positions on the test masses and the same nominal loss as LLO. 
We assume that these curves represent the likely extremum of PRG behavior for a single point absorber on an optic. From the range, it is reasonable to expect large differences between hypothetical individual interferometer configurations.

%% file: fitTable.tex

\begin{center}
 \begin{tabular}{c | c | c} 
  {\bf Interferometer } & {\bf $\mathcal{L}_\mathrm{nom}$ (ppm) } & {\bf $b$ (ppm) } \\ [0.5ex] 
 \hline \hline 
 LHO & 68.0 & 1.0  \\ 
 \hline 
 LLO (pre-realign) & 60.2 & 6.4  \\ 
 \hline
 LLO (post-realign) & 60.7 & 1.3  \\ 
 \hline
\end{tabular}
 \end{center}

%% file: TimeConstantOfPRG.tex
\begin{figure}
  \centering
  \includegraphics[width=\columnwidth]{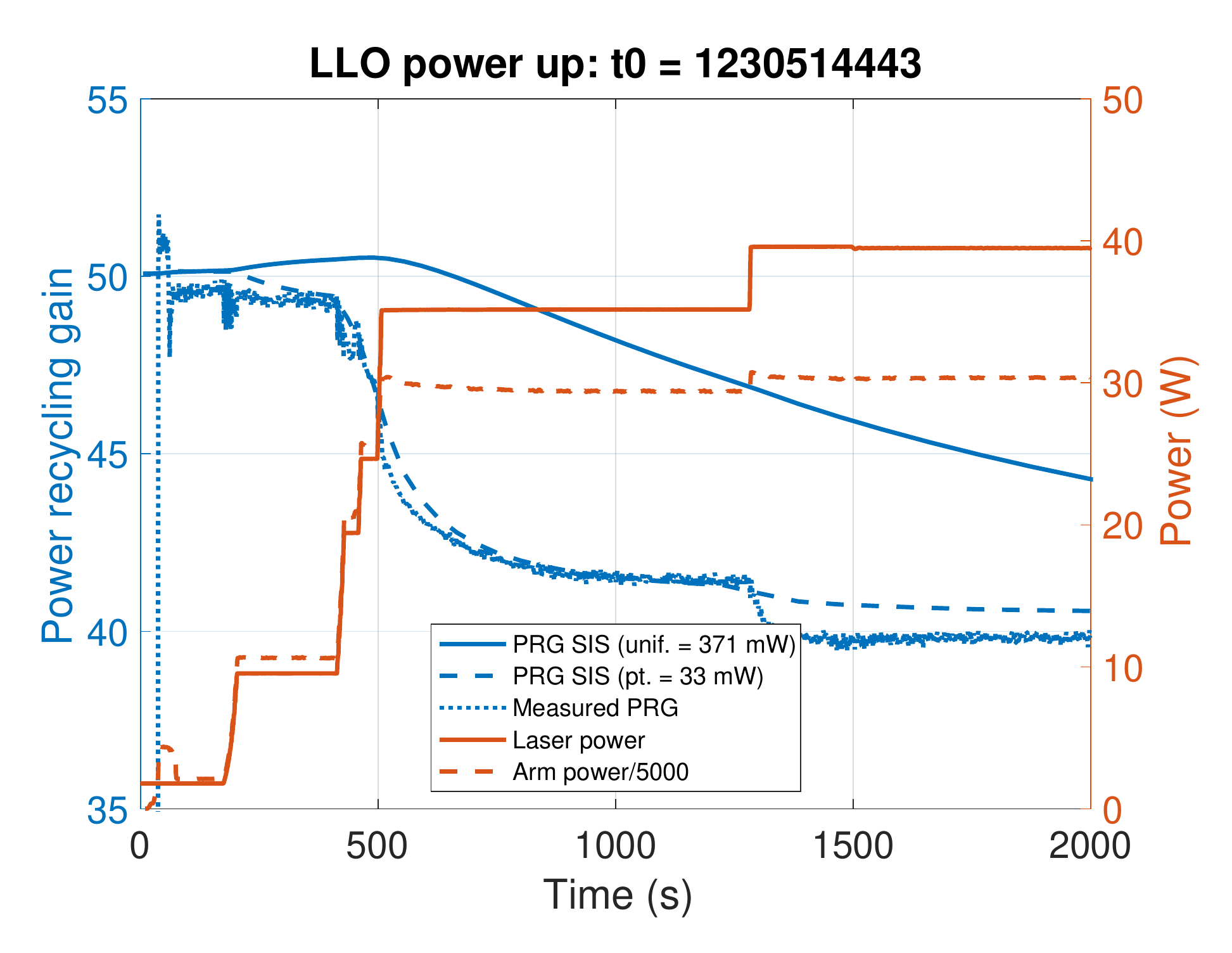}
  \caption{Time evolution of power recycling gain during a power-up at LLO. Laser power and (scaled) arm power are indicated by  the right-hand axis. Power recycling gain is shown on the left-hand axis. Also shown are two SIS models of the power-recycling gain assuming surface deformation on a single ETM from, uniform (unif.) and a point absorber (pt), respectively. The time-evolution of the surface deformation was modeled in COMSOL and scaled in amplitude to yield a PRG of 40 at $t = \infty$.}
  \label{fig:time_constants}
\end{figure}

We measured a rapid drop in the power recycling gain  with a time scale on the order of 200s. This is illustrated in Figure \ref{fig:time_constants}.

Using a combination of a COMSOL finite-element thermal model of point or uniform absorption induced surface deformation on a test mass and a numerical FFT model of the full interferometer (the Static Interferometer Simulation (SIS) \footnote{{https://labcit.ligo.caltech.edu/~hiro/SIS/}}), we simulated the time evolution of PRG as the interferometer was powered up for two cases. 
In one case we assumed uniform absorption and in the other, we assumed a point absorber located approximately \SI{20}{\milli\meter} from the center of the optic.


The LLO arm power, $P_{L}(t),$ is shown in Figure \ref{fig:time_constants} (dashed red curve). 
In two time-dependent finite-element analyses, we modeled surface deformations, $\Delta s_{i}(x,y,t)$,  from uniform absorption and from a point absorber as a function of time, scaling the absorbed power by $P_{L}(t)$ in the two cases.

The magnitude of these surface deformations were scaled such that, when added to an ETM in SIS, the steady-state power recycling gain was equal to 40, the same value measured in the interferometer (dotted blue curve in Figure \ref{fig:time_constants}). 
Finally, the SIS model was run on all times from $t=0$ to steady-state (\SI{7200}{\second}) and  predicted the power-recycling gain.

As the results show, the rapid drop in PRG is predicted by the point-absorber SIS model. This is particularly striking when  compared to the uniform absorption SIS model  that shows a time scale more than an order of magnitude larger. The smaller drop in PRG around \SI{1300}{\second} is thought to be due to slight misalignment.

Consistent with the time-scale analysis in Appendix \ref{sec:HOM_scattering_tau}, this result is highly suggestive that the majority of the power-recycling gain reduction that we see can be attributed to point absorbers interacting with higher order modes rather than uniform absorption.



%% file: ITMnoise.tex

\begin{figure}[h]
    \centering
    \includegraphics[width=\columnwidth]{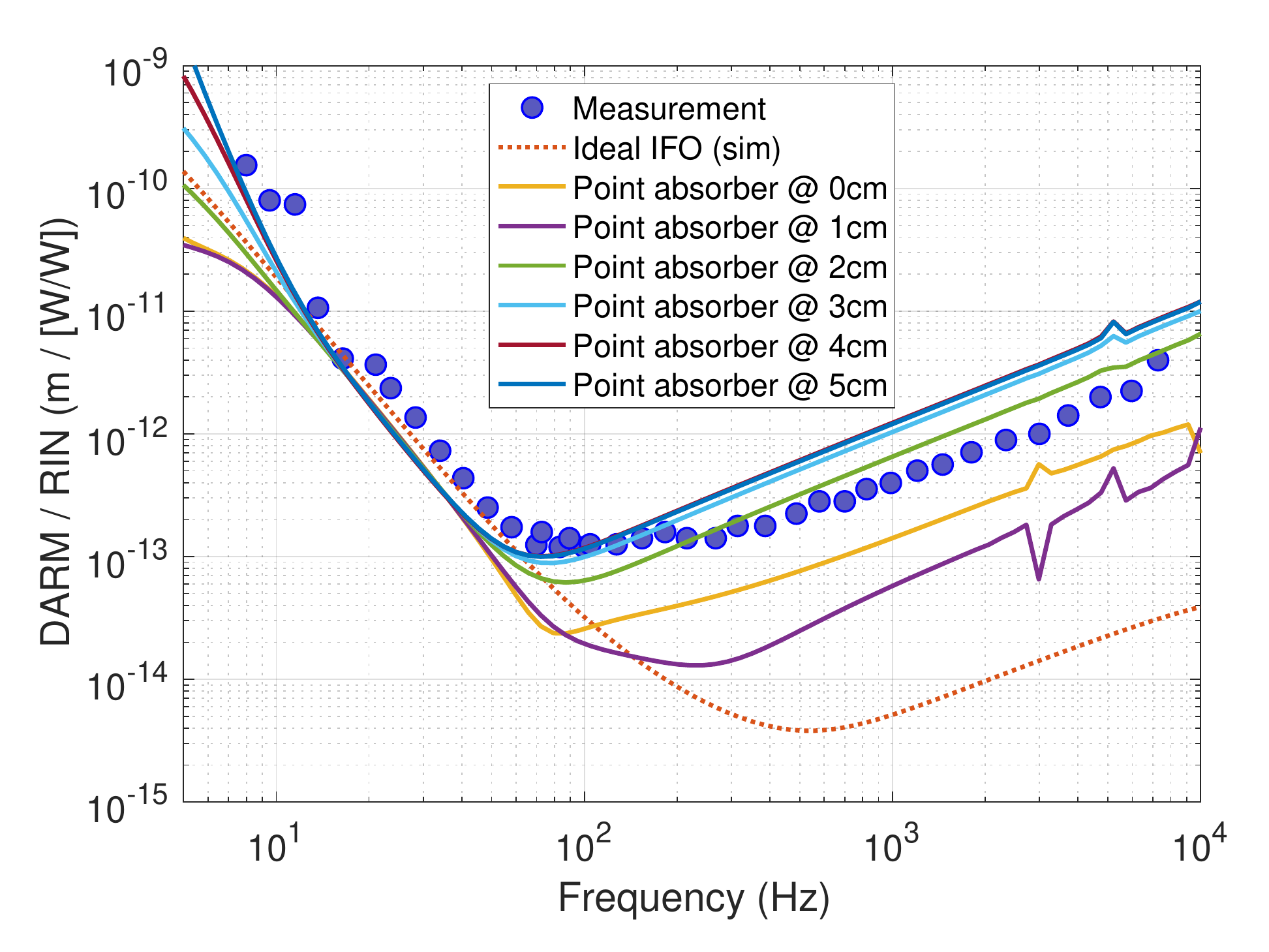}
    \caption{RIN coupling models versus measurements from the LHO interferometer. The "ideal IFO" trace is for a perfectly symmetric IFO, therefore the coupling is due to radiation pressure at low frequency and sidebands RIN at high frequency. The other traces have the expected spherical lenses on the ITM substrates, plus one point absorber map at different distances from the center. Simulations were created using the MIST simulation tool \cite{Vajente_2013}. }
    \label{fig:RIN_coupling}
\end{figure}

We measured the coupling of the input laser relative intensity noise (RIN)  to differential arm motion (DARM) in aLIGO. At frequencies above  \SI{500}{\hertz}, the coupling was approximately two orders of magnitude higher than expected from simulations of an ideal interferometer, as shown in Figure \ref{fig:RIN_coupling}. However,  simulations including ITM thermal lenses from point absorbers modeled at a variety of different radii showed significant increase in high frequency RIN coupling, sufficient to explain the excess coupling observed in the interferometer. 

Uncertainty in point absorber position creates a large uncertainty in the exact coupling coefficient. Hence, the model can only reproduce the qualitative behavior of the observed RIN coupling. Noise budget measurements reported in Buikema et al.\cite{O3CommissioningPRD} show the absolute laser intensity contribution to overall noise is not yet a limiting noise source. Below \SI{2}{\kilo\hertz}, it is at least   $10\times$ lower  than the differential arm motion in both LIGO interferometers (Figure [2] in Buikema et al.\cite{O3CommissioningPRD} explicitly shows this). However, the coupling, and hence the noise contribution, is expected to increase as arm cavity power increases in future.

%% file: FutureIFOs.tex
\label{sec:future}

\begin{figure}
  \centering
  \includegraphics[width=\columnwidth]{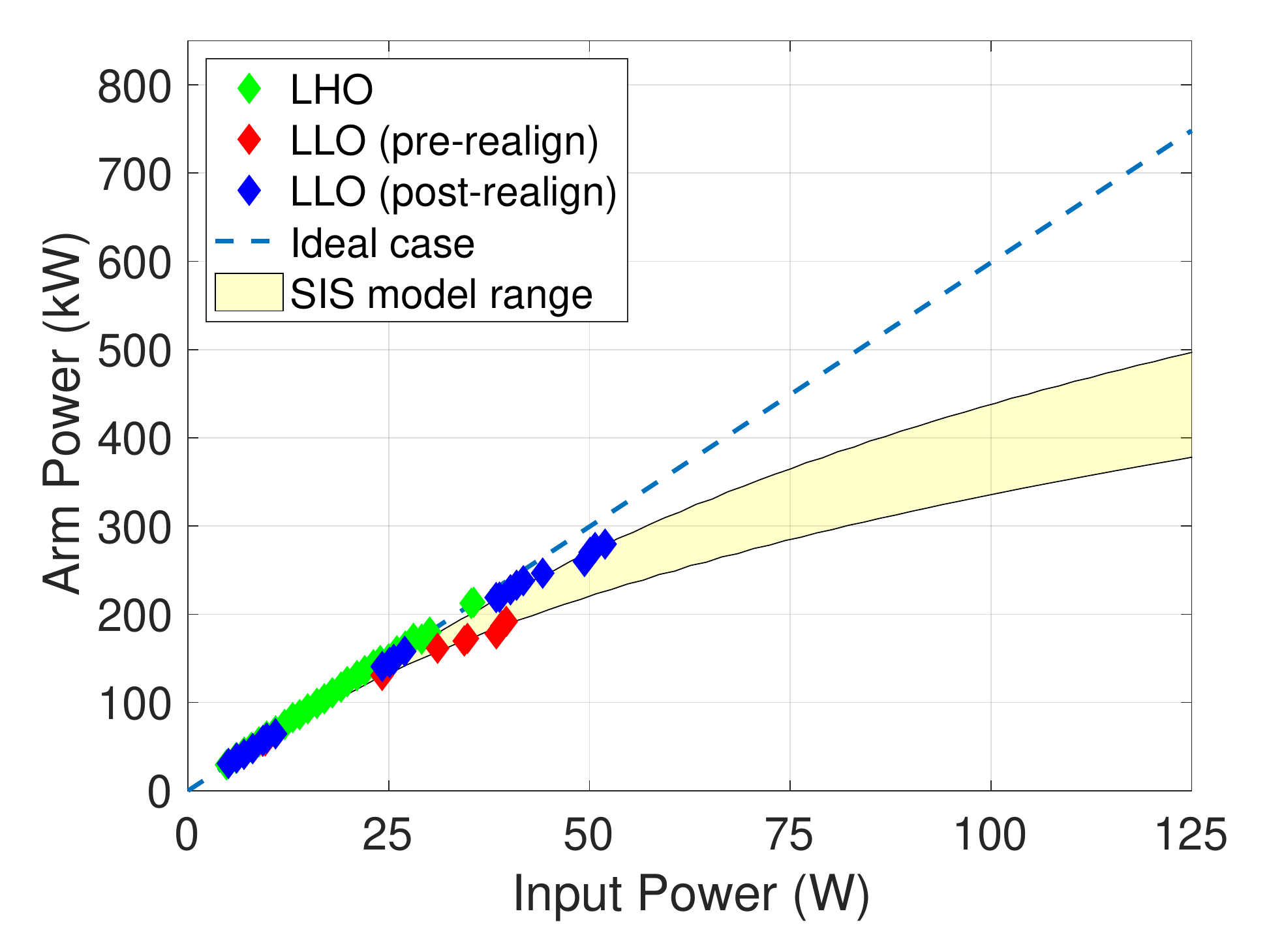}
  \caption{Arm power versus input laser power. The dashed line shows the case of constant optical gain. The data points show the measurements from the LIGO sites. The yellow range shows predictions from the SIS model assuming uniform absorption and a variety of point absorbers on the optic consistent. As the effect of the point absorber is decreased, the uniform absorption becomes the limiting factor preventing ideal build-up of arm power. }
  \label{fig:BS_power}
\end{figure}

It is clear that point absorbers adversely affect the performance of the aLIGO interferometers and that this effect increases as input power is increased. 
In Figure  \ref{fig:BS_power} we plot the projected  arm power versus input power assuming the most optimistic scenario of solely a power-dependent reduction in power-recycling gain. 
This projection corresponds to the three different scenarios in Figure \ref{fig:PRG_v_power} (O3 LHO (green), O3 LLO pre-realignment (red) and O3  LLO post-realignment (blue)). 
Briefly, all scenarios predict a significant reduction in the maximum stored power (and, hence, performance) for aLIGO.

Despite the improvement in PRG attained by repositioning the beam on the ETM, the SIS model predicts a minimum 33\% deficit in arm power build-up at nominal operating power of \DataVal{MaxInputPower}{}. Assuming no other adverse effects to interferometer operation, we would expect a corresponding increase in the shot noise floor of 15\%$\--$20\% above the nominal Advanced LIGO noise floor. 

Mitigating the adverse effects of point absorbers can be accomplished by addressing all the features identified in Section \ref{sec:prg_generalized_discussion} and expressed by the amplitude and gain terms in Equation \ref{eqn:amp_gain}. 

Amplitude reduction strategies seek to limit or eliminate the single bounce scattering term, $a_{00|mn}$, in Equation \ref{eqn:amp_gain}. The first, and ideal, scenario is to eliminate absorbers at the source when they are introduced into the coating: either through modifications to the coating process or active elimination after coating. This is being actively researched within the LIGO Scientific Collaboration. Beyond that, the simplest method for partial reduction of amplitude scattering in-situ is re-positioning the interferometer beam, as demonstrated at LLO. Additionally, scattering to specific HOM can be reduced by actuating on those modes using, for example, a high-spatial frequency corrector \cite{CHRAAC:2013}. Finally,  surface deformation (and therefore amplitude scattering) are minimized for different optic materials; for example, the proposed use of cryogenic silicon in future interferometers \cite{Voyager_2020, NEMO:2020} has the benefit of an, effectively, zero coefficient of thermal expansion.

Resonant enhancement/suppression techniques seek to reduce the resonant gain term, $g_{mn}$, in Equation \ref{eqn:amp_gain}. Future interferometers are free modify the overall cavity design and g-factor with respect to HOM spacing to minimize the effect from point absorbers. A more targeted option, actively being explored for Advanced LIGO, is to deliberately polish surface errors into the edges of the optic, such as  those illustrated in Figure \ref{fig:ETM_cross_section},   creating a wider HOM free region  around the fundamental mode resonance \cite{Hiro2020}. In addition, conceptual work has begun on active front surface thermal actuators that could produce such surface errors dynamically - thus allowing for in-situ fine-tuning of the HOM resonances around the fundamental mode.

%% file: conclusion.tex
\label{sec:conclusion}

We have shown that point-like absorbers present a serious impediment to obtaining full operating power in Advanced LIGO, primarily due to increased arm losses as absorbed power increases. We reiterate that  that errors in the mirror surface profile, mirror aperture effects and higher order mode behavior matter are crucial to understanding considering these losses.  Future gravitational wave detectors call for stored power approximately an order of magnitude or more higher than O3 \cite{Miao_HF:2018, Martynov:2019_PRD, Page_HF_GW:2020, ET2010, Reitze2019Cosmic}, with similar laser intensities to full-power Advanced LIGO, so  the existence of point absorbers and their interaction with the optical fields of those detectors is of significant importance.

We have explored the physics of cavity-optic-deformation to explain the observed reduction in interferometer performance whilst, simultaneously, highlighting those elements that are most crucial to mitigating the effects of point absorbers in Advanced LIGO and future gravitational wave interferometers.  

%% file: aLIGOOperation.tex
\label{sec:aLIGOOperation}


To understand the observed interferometer behavior, it is helpful to  provide an overview of how the aLIGO interferometer works. 
This section contains a description of the ideal operation of the Advanced LIGO interferometer. Interested readers can find   more details   in other references \cite{aLIGO_Overview, aLIGO_ASC_Barsotti_2010, aLIGO_Staley_2014, aLIGO_Martynov_2016, aLIGO_LSC_2016, aLIGO_O2_Driggers}.

\subsection{Detector sensitivity}

As shown in Figure \ref{fig:IFO_schematic}, Advanced LIGO is a suspended Michelson interferometer,  modified to maximize sensitivity to GW with a bandwidth of approximately \SI{20}{\hertz} \-- \SI{2}{\kilo\hertz}. All modifications are designed to either amplify the signal strength, reduce the contribution of different noise sources or improve robustness and reliability.

A passing gravitational waves of amplitude, $h$, induces  a strain in spacetime, lengthening one interferometer arm and shortening the other by an amount, $\Delta L$, where 

\begin{equation}
    \Delta L = h\, L \label{eqn:DeltaL}
\end{equation}

\noindent and $L$ is the interferometer arm length. This imposes phase fluctuations on the laser beams present in those arms (GW audio sidebands) that are converted,  by the beamsplitter,   into  intensity fluctuations  on the output side of the Michelson interferometer. 

As Equation \ref{eqn:DeltaL} shows, the resulting change in arm length (and, hence, the signal strength) is proportional to the arm length. To exploit this, the aLIGO arms are  \SI{4}{\kilo\meter} long, substantially amplifying the signal for a given strain relative to a short interferometer. 
The arms contain resonant Fabry-Perot cavities that  increase the average number of round-trips  photons experience in the arms, and the phase they accumulate, by the optical gain of the arms, $G_A$. For the Fabry-Perot arm cavity, since the input and end mirror transmissions are very low ($T_i = $ \DataVal{ITMTrans}{} and $T_e = 5$ppm, respectively), the following approximation holds:

\begin{equation}
    G_A \approx  \frac{4}{T_i}
\end{equation}

\noindent resulting in a nominal phase amplification factor of  approximately \DataVal{ArmCavityFinesse}{}  for aLIGO. (The optical gain is also approximated by $2\,F/\pi$, where $F$ is the cavity finesse, approximately 440).

Ideally, the Fabry-Perot arms resonate the fundamental Gaussian spatial mode (sometimes referred to as TEM00) and suppress higher order spatial modes (HOM). Higher order modes can be described in terms of different bases, for example, Hermite-Gauss (HG) or Laguerre-Gauss (LG) \cite{Siegman:Lasers}. 

In contrast to the limited ways to amplify the signal, there are a large number of independent noise sources in the interferometer. The majority of noise sources limit the sensitivity  below approximately \SI{100}{\hertz} \cite{Martynov_noise:15, aLIGO_O2_Driggers}. 
However, at high frequencies, an ideal detector is limited only by fundamental quantum sensing noise (shot noise). The Signal-to-Noise Ratio (SNR) from shot noise is  proportional the square root of the power incident on the input (symmetric) side beamsplitter. Thus, one of the the simplest ways to improve the detector sensitivity is to increase  the input laser power (up to \SI{125}{\watt} for aLIGO).

The  incident power on the beamsplitter is further  increased by adding a power recycling mirror (PRM), see Figure \ref{fig:IFO_schematic}, to return power coming  out of the interferometer back into it. 
This mirror forms another resonant cavity, named the power recycling cavity. The coupled power-recycling and common arm cavities are referred to as the common coupled  cavity. 

For the power recycling to maintain a high optical gain, the interferometer is held close to a dark fringe on the output (anti-symmetric) side of the beamsplitter, sending nearly 100\% of the optical power on the beamsplitter back to the PRM. 
The differential arm length is held slightly off minimum (close to dark fringe) such the static amount of power at the AS port is held constant to serve as a local oscillator for DC (homodyne) readout \cite{Hild_2009}. 

The coupled power-recycling and arm cavities  increase  the optical gain such that the power stored in the PRC and a single arm are approximately $50\times$ and $6000\times$ the input power, respectively (in fact, these numbers vary in the real interferometer depending on losses in the arms and power-recycling cavities). 

The storage time of photons the arm cavity is similarly increased. Intensity noise present on the light resonant in the cavity is filtered above $1/\tau_{CC}$, where $\tau_{CC}$ is the coupled cavity storage time, corresponding to a pole at a frequency of approximately \DataVal{CCpole}{} \cite{Martynov_noise:15}. This filtering is essential to suppress the coupling of relative intensity noise (RIN) from the input laser to the GW channel. 

As shown in Figure \ref{fig:IFO_schematic}, on the output side of the interferometer a  recycling mirror (SRM) is present, forming a resonant cavity with the test masses called the signal recycling cavity (SRC).
The SRC is held on anti-resonance, lowering optical gain and storage time of difference signals in order to increase the signal bandwidth of the interferometer (known as resonant signal extraction \cite{RSE:2003}).
The combination of the signal-recycling and arm cavities is called the differential coupled  cavity. 

Finally, the output of the interferometer is spatially filtered through an output mode cleaner (OMC) to remove   higher order spatial modes that carry no GW signal, and strip off any residual radio-frequency (RF) control sidebands from the laser, described below.  At the output of the OMC the GW signal sidebands beat with the static offset induced by the arm differential offset to provide a power fluctuation that is detected by a pair of photodiodes.

\subsection{Power-recycling gain}

The power recycling gain, $G_P$, defined as the ratio of the stored power inside the power recycling cavity to the input laser power incident on the PRM, is a function of the reflectivities of the power-recycling mirror, input test masses and end test masses and also a function of the losses in the arms and the PRC. 
The amplitude reflectivities are denoted in lower-case, $r_a$, and corresponding power reflectivities are denoted in upper-case, $R_a = r_a^2$. The values are given in Table \ref{tab:values} in Appendix \ref{sec:table}.


For this analysis, it is convenient to consider the two arm cavities as an aggregate (common) arm cavity and the power-recycling cavity as a three mirror cavity made of the PRM, an average Fabry-Perot arm. The power recycling gain is given by:

\begin{equation}
    G_P = \left( \frac{t_p}{1-r_p\,r_{FP}\, \left(1-\mathcal{L}_P/2\right)} \right)^2 
    \label{eqn:GP0}
\end{equation}

\noindent when the PRC is on resonance \cite{AndoThesis}. The amplitude reflectivity of the average arm cavity,  $r_{FP}$ \cite{AndoThesis}, is given by:
\begin{eqnarray}
    r_{FP} & = & \frac{-r_i + r_e^*}{1 - r_i\, r_e^*} \\
     & \approx & 1 - 2 \frac{T_e}{T_i} - \frac{G_A\, \mathcal{L}_A}{2} \label{eqn:rFPs}
\end{eqnarray}

\noindent when the arm cavities are on resonance and $T_e,T_i<<1$. 
Combining equations \ref{eqn:GP0} and \ref{eqn:rFPs} yields an expanded form of the power-recycling gain,

\begin{equation}
    G_P \approx \left( \frac{t_p}{1-r_p\,  \left(1 - (G_A\,T_e - G_A \, \mathcal{L}_A -\mathcal{L}_P)/2\right)} \right)^2  \label{eqn:GP_gen}
\end{equation}

\noindent in which the arm loss contribution is greater than the recycling cavity loss by a factor of $G_A$, approximately \DataVal{ArmCavityFinesse}{} for aLIGO. Thus, the power-recycling gain can be used as a proxy for losses, particularly arm losses, in the interferometer.

\subsection{Feedback control}
\label{sec:aLIGOOperation_2}
In order for the interferometer to function in low noise, all of the longitudinal degrees of freedom (DOF) in the interferometer 
must be sensed and locked onto resonance for the main laser field (the GW carrier). 
This is achieved by using the Pound-Drever Hall locking technique that adds pairs of radio frequency sidebands  onto the main laser \cite{PDH_1983} at $\pm$\SI{9}{\mega\hertz} and $\pm$\SI{45}{\mega\hertz}. 
As illustrated in Figure \ref{fig:IFO_schematic}, both pairs of  sidebands are anti-resonant in the Fabry-Perot arms, resonant in the PRC. The \SI{45}{\mega\hertz} sidebands are resonant in the SRC. Full details can be found in \cite{Martynov_noise:15}.

Since sidebands and carrier fields have different frequencies, their propagation inside the meters- or km-long interferometer cavities produce different phase shifts. The net result is that carrier and sideband fields have different resonant conditions in the various interferometer cavities. In particular, while the carrier is tuned to be resonant simultaneously in the power recycling cavity and in the Fabry-Perot arm cavities, the sideband frequency is chosen so that the corresponding fields are close to anti-resonance in the arms and resonant in the power recycling cavity. In this way the sidebands provide a phase reference which is independent of the arm cavity motions, and suitable error signals can be extracted by demodulating the signals produced by fast photodiodes at the same modulation frequency \cite{Martynov_noise:15}. 

Thus linear combinations of this beating of these sidebands and carrier, measured on RF photodiodes at different locations and in different modulation quadrature yield enough control signals to uniquely measure all the desired degrees of freedom and, when fed back into a control system, keep the residual longitudinal motions less than the required levels. 
The noise in the control system must be kept very low lest the feedback loop inject it back into the interferometer at a level which then limits the detector sensitivity. 

In the interests of brevity, the above description of the feedback control system is deliberately terse. A full description of the control is beyond the scope of this text, references are provided \cite{HallThesis_2017, MartynovThesis}. However, one element key to the following discussion concerns the differential arm motion (DARM): the degree of freedom that encodes the gravitational wave signal. For the subsequent discussion, it is important to  know that the feedback control system is set up to maintain a constant DC power (\SI{20}{\milli\watt}) on the output photodiodes by feeding the DARM error signal back to the differential arm length. 


In summary, Advanced LIGO achieves good sensitivity at high frequencies by having large amounts of stored power in those cavities and incident on the the test masses while relying on a low-noise control system to keep the optical cavities on resonance  within minimal residual longitudinal motions.

%% file: optics_and_pt.tex
\label{sec:Optics}

\begin{figure}[h]
  \centering
  \includegraphics[width=11cm]{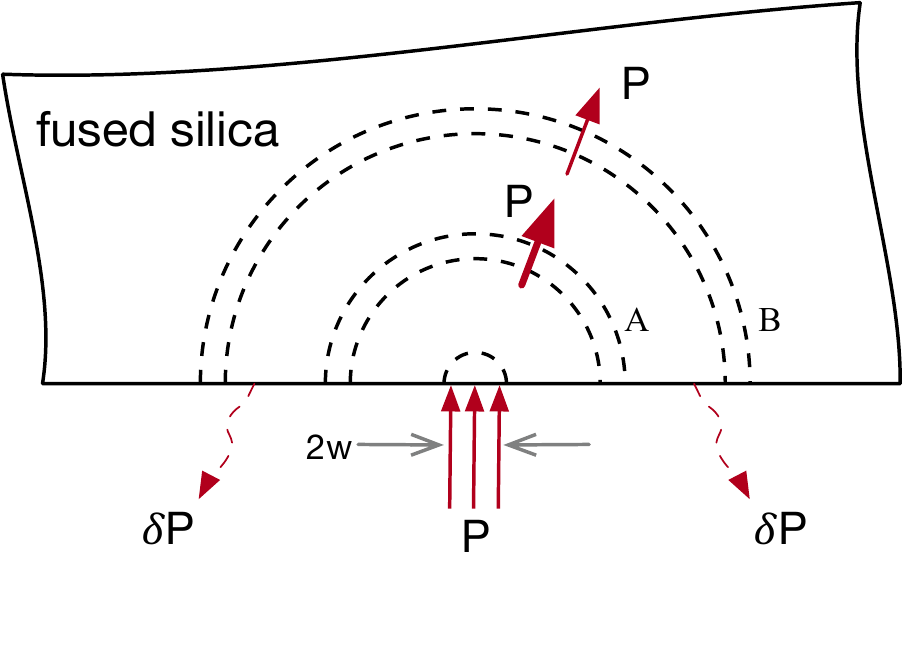}
  \caption{Power flowing into a point-like region. The radiated power is negligible compared to the conduction. Hence, all power is effectively transferred by conduction. The far-field gradient is only determine by the absorbed power and thermal conductivity.}
  \label{fig:point_absorber}
\end{figure}

Consider a point-like absorber, absorbing within a diameter $2\, \omega$, on the surface of an optic that is exposed to optical power, as illustrated in Figure \ref{fig:point_absorber}. We solve for the surface deformation inside and outside the absorbing region.

In the specific case that the absorbing region is on the surface of an optic within a LIGO arm cavity, the absorbed power, $P_\mathrm{abs}$, is determined by the local intensity of the illuminating beam multiplied by the absorbing area:

\begin{equation}
    P_\mathrm{abs} = \left(\pi\, \omega^2\right) \, \frac{2\, P_\mathrm{A}}{\pi \, \mathrm{w}^2} \, \exp\left[-\left(\frac{\mathrm{r_c}}{\mathrm{w}}\right)^2 \right] \label{eqn:power_abs}
\end{equation}

\noindent where $P_\mathrm{A}$ is the laser power stored in the arm,  $\mathrm{w}$ is the Gaussian beam radius and $\mathrm{r_c}$ is the position of the absorbing region relative to the center of the Gaussian beam. For simplicity, we have assumed that the  region is 100\% absorbing.

\subsection{Small and large scale surface deformation and time scales}

\subsubsection{Inside the absorbing region}
Winkler et al. solved for the approximate surface deformation, $\Delta s_\omega$, on the surface of an optic, within the absorbing region \cite{PhysRevA.44.7022}, finding

\begin{equation}
    \Delta s_\omega = -\frac{\alpha}{4\,\pi\,\kappa}P_{\mathrm{abs}}
\end{equation}

\noindent where $\alpha$ is the coefficient of thermal expansion, $\kappa$ is the thermal conductivity and $P_{\mathrm{abs}}$ is the power absorbed within the absorbing region. The negative sign is purely convention based upon the deformation expanding {\it away} from a mirror. Expressed as a quadratic function of position:

\begin{equation}
    \Delta s = -\frac{\alpha}{4\,\pi\,\kappa}P_{\mathrm{abs}} \left(\frac{r}{\omega}\right)^2. \label{eqn:WinklerDS}
\end{equation}

\subsubsection{Outside the absorbing region}

We solve for the approximate surface deformation for a half-infinite cylinder (i.e. assuming radial and longitudinal boundaries are far away). We assume only conduction and ignore power radiated from the front surface of the optic - given that the power radiated from one beam radii is of the order $2\%-3\%$ of the total power radiated by the optic.

Ignoring boundary conditions and considering the region outside the radius of the point absorber, the temperature distribution is solely governed by the equation of thermal conductivity:

\begin{eqnarray}
    P_{\mathrm{abs}} & = & -\kappa\,A\,\overrightarrow{\nabla}T
\end{eqnarray}

\noindent where $A = 2\,\pi \, r_S^2$, the area of a hemispherical shell in the optic, where $r_S$ is the spherical radial coordinate ($\sqrt{x^2+y^2+z^2}$). Solving in spherical coordinates, the temperature profile is given by 

\begin{eqnarray}
    T (r) & =&  \frac{P_{\mathrm{abs}}}{2\, \pi\, \kappa\,r_S} \\
    & = & \frac{P_{\mathrm{abs}}}{2\, \pi\, \kappa\,\sqrt{r^2 + z^2}}
\end{eqnarray}

\noindent where $r=\sqrt{x^2+y^2}$ and $z$ are the polar radial and longitudinal coordinates, respectively. The surface deformation is approximated by the coefficient of thermal expansion, $\alpha$, multiplied by the integral of temperature field:


\begin{eqnarray}
    \Delta s &= & \alpha\,  \int_0^{h}T(r)\,\mathrm{d}z
\end{eqnarray}

\noindent where $h$ is the thickness of the optic. Combining this with equation \ref{eqn:WinklerDS} and solving yields a generalized approximation for the surface deformation on the optic:


\begin{equation}
    \Delta s \approx \frac{\alpha\,P_{\mathrm{abs}} }{2\,\pi\,\kappa} \, f(r) \label{eqn:outside_abs}
\end{equation}

\noindent where

\begin{equation}
f(r) = 
\begin{cases}
  -\frac{1}{2}\,\left(\frac{r}{\omega}\right)^2 &  r \leq \omega  \\
 c_0+ \log\left[ \frac{h + |(r,h)|}{r}\right] &  r > \omega  
\end{cases}
\label{eqn:fr_defn}
\end{equation}


\noindent where $c_0$ is a constant offset such that the two parts have the same value at $r = \omega$:

\begin{equation}
    c_0 = -\frac{1}{2} - \log\left[ \frac{h + |(\omega,h)|}{\omega}\right]
\end{equation}

\noindent and where

\begin{equation}
|(r,h)| = \sqrt{r^2 + h^2}.
\end{equation}

\begin{figure}[h]
  \centering
  \includegraphics[width=11cm]{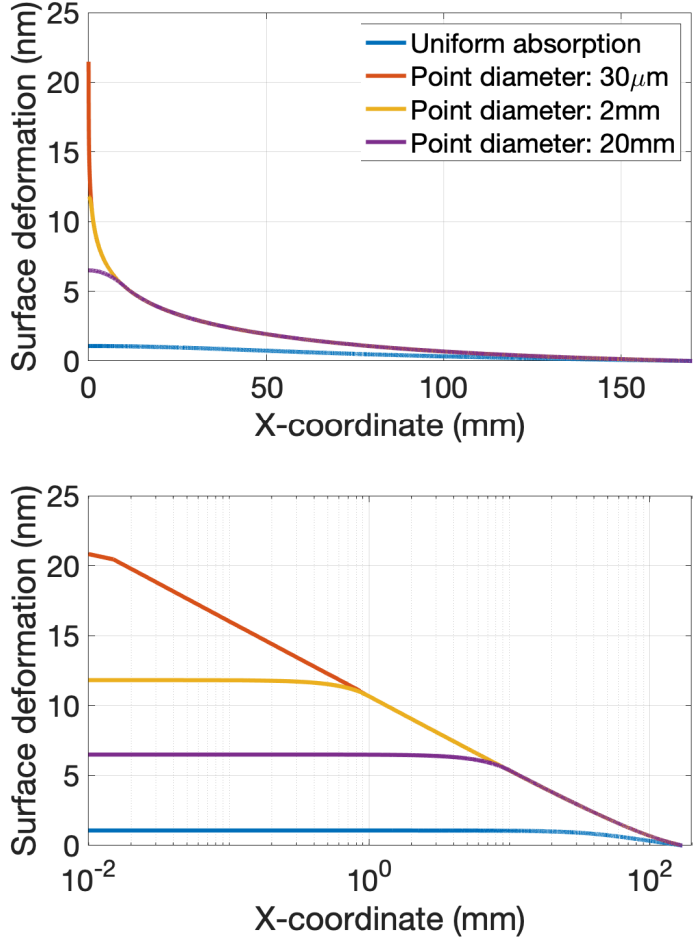}
  \caption{Cross-section of modeled surface deformation from uniform and point absorbers shown in linear (upper) and semilog (lower) plots. The red, yellow and purple curves 
  show surface deformation from point absorbers of different physical size but with each absorbing \SI{40}{\milli\watt}. Uniform absorption of \SI{10}{\milli\watt} (blue) is shown for comparison. All point absorber cases show identical surface deformation outside the absorbing region.}
  \label{fig:model_SD}
\end{figure}

Note that the surface deformation has been referenced to the center of the point absorber such that the surface deformation is zero at that point. Equation \ref{eqn:outside_abs} shows that outside of the absorbing region, the large scale surface deformation depends only on the absorbed power and not on the size or distribution of the absorbing region. This is illustrated by a finite-element model of surface deformation from point absorbers of different size but fixed absolute absorbed power, shown in Figure \ref{fig:model_SD}. The surface deformation from uniform absorption is shown for comparison. 

\subsubsection{Time scale}

Finally, the time scale for a temperature distribution to form is governed by the heat equation: 

\begin{equation}
    \frac{\partial T}{\partial t} = D \, \left( \frac{\partial^2 T}{\partial x^2} + \frac{\partial^2 T}{\partial y^2} + \frac{\partial^2 T}{\partial z^2}  \right) 
\end{equation}

\noindent where $D$ is the thermal diffusivity of the mirror, given by:

\begin{equation}
    D = \frac{\kappa}{\rho\, c}
\end{equation}

\noindent where $\rho$ is the density, $c$ is the specific heat capacity and $\kappa$ is the thermal conductivity. A cursory examination of the heat equation shows that the time-scale for the temperature distribution to form over a given spatial scale will be proportional to that spatial scale squared. For example, for a given spatial scale, $r_0$, the time constant for the temperature distribution (and thus surface deformation) to form is:

\begin{equation}
    \tau_{r_0} \propto \frac{r_0^2}{D}.
\end{equation}

Thus, time-evolution of surface deformation is an effective discriminator for different spatial scales. We  return to this idea in the next appendix where we evaluate  time evolution for specific spatial distribution of higher order optical modes. 

%% file: single_bounce_scattering.tex
\label{sec:single-bounce}
 
Consider the immediate effect of reflection of a fundamental Gaussian (TEM00) optical field, of power $P$, from a deformed mirror surface -  hereafter referred to as the single-bounce reflection off the optical surface. The  distribution of  this field is given by:

\begin{equation}
    E_{00} = \left(\frac{2\, P}{\pi \, \mathrm{w}^2} \right)^{1/2} \exp\left[-\left(\frac{r}{\mathrm{w}}\right)^2 \right]
\end{equation}

Specifically, we consider the magnitude of scattering into higher order optical modes and the characteristic time scale for such scattering. Note that in the following discussion, we justify ignoring the curvature of the mirror and the curvature of the incident Gaussian beam by assuming that these are identical.
 

\subsection{Scattering as a function of point absorber  size and position}
\label{sec:single-bounce_size}
 
Upon reflection from a mirror with  surface deformation, $\Delta s(x,y)$, the scattering from the fundamental mode to the $mn$-th HOM is determined by the overlap integral of the TEM00 mode with the higher order mode, $E_{mn}$, plus the phase error, $i\,k\,\Delta s$, multiplied by a factor of 2 to account for the double-pass upon reflection. Thus, the complex amplitude coefficient is:

\begin{eqnarray}
    a_{00|mn} & = &  \iint \mathrm{E_{mn}} e^{\mathrm{i}\, k\, (2\,\Delta s)} \mathrm{E_{00}} \,\mathrm{d}x\, \mathrm{d}y \\
    & \approx & \mathrm{i}\,2\,k\, \iint \mathrm{E_{mn}} \,\Delta s\, \mathrm{E_{00}} \,\mathrm{d}x\, \mathrm{d}y \label{eqn:linear_amp_scatt}
\end{eqnarray}

\noindent where $k = 2\pi/\lambda$ and the approximation is valid for deformations that are significantly smaller than the wavelength of light. In the case of a point absorber, we combine with Equation \ref{eqn:outside_abs} and find:

\begin{equation}
    a_{00|mn}  \approx  \frac{\mathrm{i}\,2\,\alpha\, P_\mathrm{abs}}{\lambda\, \kappa} \iint \mathrm{E_{mn}} \, f(r)\, \mathrm{E_{00}} \,\mathrm{d}x\, \mathrm{d}y \label{eqn:a00mm_w_fr}
\end{equation}

The amplitude scattering coefficient is proportional to the absorbed power, $P_{\mathrm{abs}}$ and, hence, the {\it power} scattering coefficient, $a_{00|mn}^2$, is proportional to the square of the absorbed power.

\begin{figure}[h]
  \centering
  \includegraphics[width=\columnwidth]{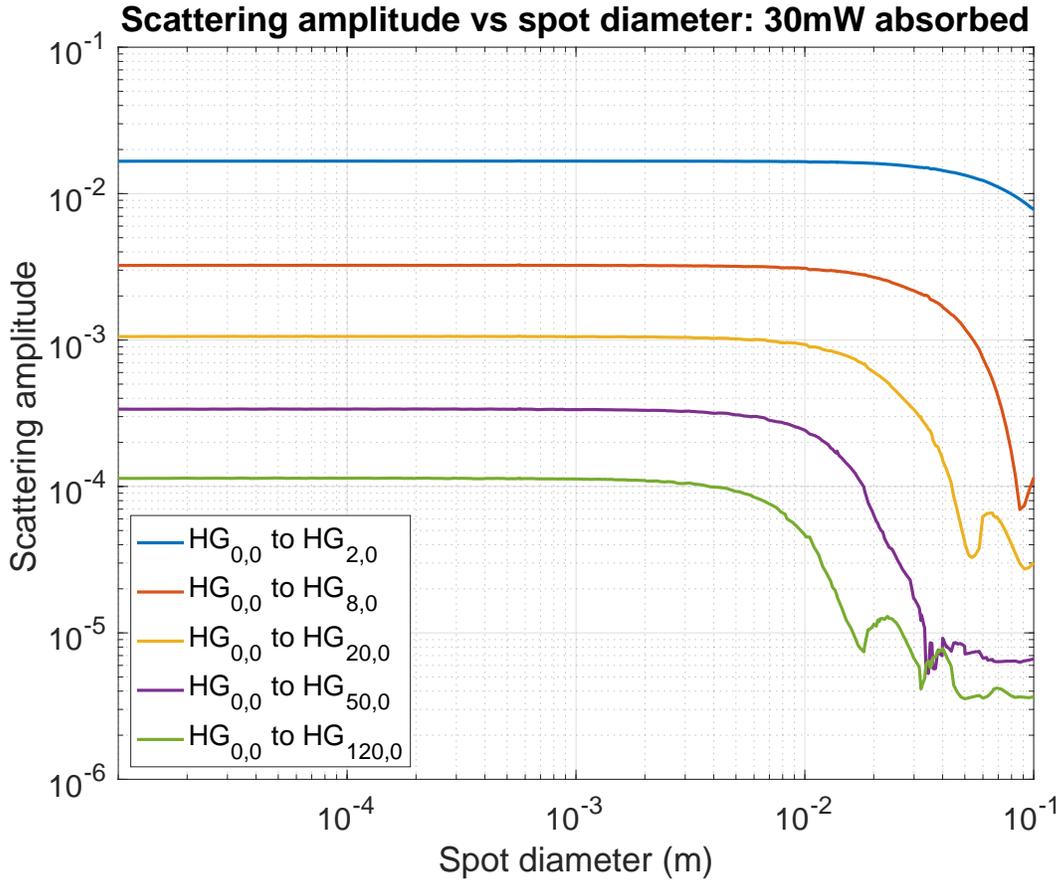}
  \caption{Single-bounce scattering amplitude into higher order modes vs point absorber diameter for a fixed absorption of 30mW in the steady-state case. For the modes shown (up to order 120), there is no dependence on spot size for points below approximately \SI{5}{\milli\meter}.}
  \label{fig:scattering_coeff}
\end{figure}

Given that  Equation \ref{eqn:outside_abs} establishes that surface deformation is largely independent of point absorber size, we expect that scattering to HOM should be similarly independent. To verify this, scattering into different HOM was determined as a function of point absorber size whilst holding the total absorber power constant. 
The results are illustrated in Figure \ref{fig:scattering_coeff} which shows the single bounce scattering into higher order modes is indeed largely invariant with point absorber size.
It is only once the absorber approaches the characteristic size of a node of the higher order mode that we begin to see variation in the scattering coefficient.



%


However, HOM have spatial dependence. Thus the scattering amplitude is a function of the point absorber position. For a point absorber displaced from the center of the Gaussian beam by an amount $\mathbf{r_c}$, the scattering coefficient is: 


\begin{eqnarray}
    a_{00|mn}(\mathbf{r_c}) & = & \mathrm{i}\,2\,k \iint \mathrm{E_{mn}} \Delta s(\mathbf{r}-\mathbf{r_c}) \mathrm{E_{00}} \,\mathrm{d}x\, \mathrm{d}y \label{eqn:scatt_posn_dependence}
\end{eqnarray}

\begin{figure}[h]
  \centering
  \includegraphics[width=\columnwidth]{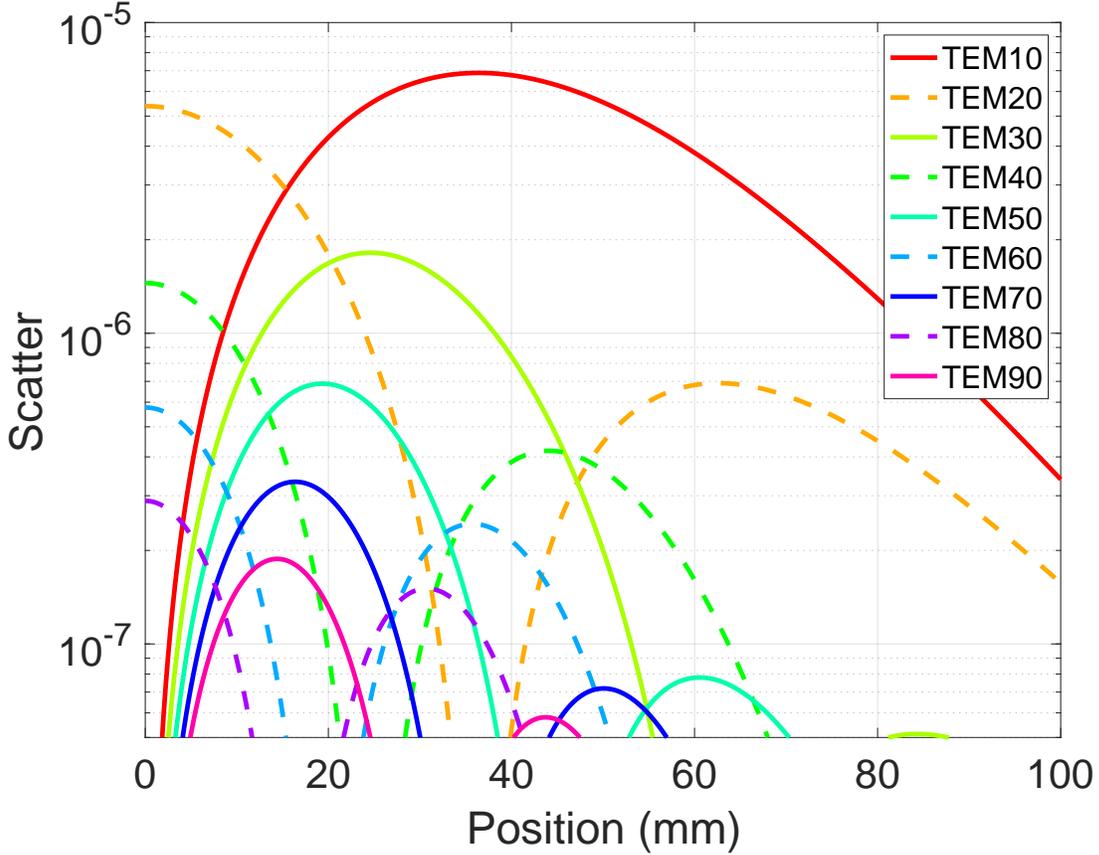}
  \caption{Scattering amplitude into Hermite-Gauss, HG$i$0, mode versus position from the center of the TEM00 field for a \SI{10}{\milli\watt} point absorber after \SI{3600}{\second} of heating. Odd modes are identified by solid lines; even modes are identified by dashed lines. }
  \label{fig:scatter_to_HOM_v_x}
\end{figure}

Scattering versus point absorber position (for a \SI{10}{\milli\watt} point absorber after \SI{3600}{\second}) is illustrated in Figure \ref{fig:scatter_to_HOM_v_x}. The scattering to even modes is maximized when the point absorber is in the center of the TEM00 field, whilst the scattering to the odd modes have zero scatter amplitude there. The scattering to odd modes is maximized at different radii with a clustering between \SI{15}{\milli\meter} and \SI{25}{\milli\meter}.






\subsection{Time scale for scattering}
\label{sec:HOM_scattering_tau}

\begin{figure}[h]
  \centering
  \includegraphics[width=11cm]{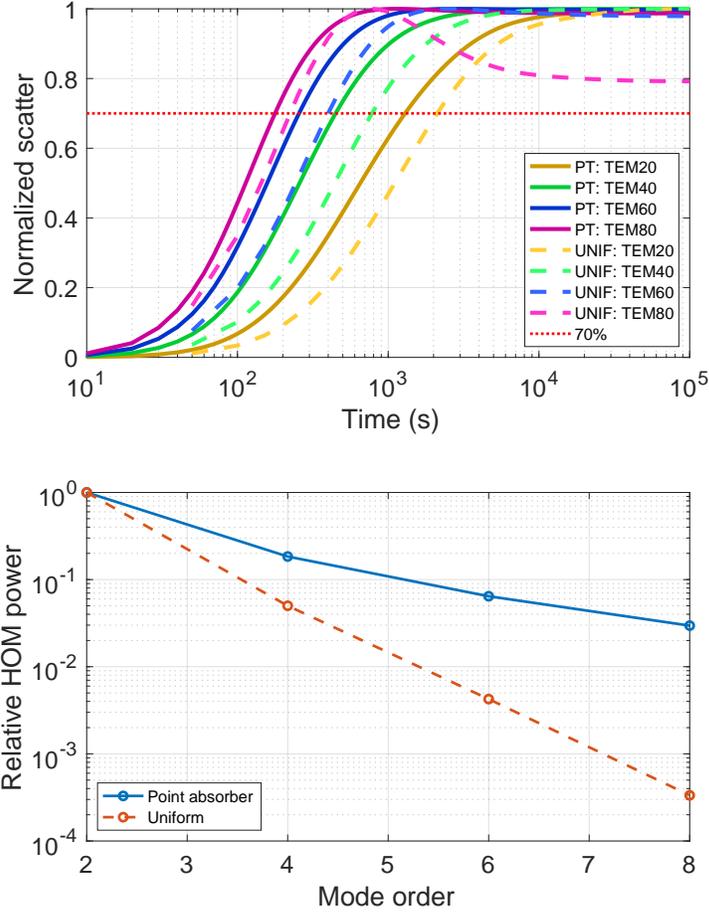}
  \caption{Top: Normalized power scattered from the incident TEM00 field into even HOM versus time for a point absorber at the center of the TEM00 mode (solid) and for uniform absorption (dashed). Bottom: power scattered into HOMs, relative to 2nd order mode [TEM20], for a point absorber (solid) and uniform absorption (dashed). }
  \label{fig:scatter_to_HOM_v_t}
\end{figure}

The scattering from the fundamental mode into even HOMs is shown in Figure \ref{fig:scatter_to_HOM_v_t}, both for a point absorber and for uniform absorption. 
The upper panel shows the {\it power} scattered into  HOM versus time, normalized to the maximum scattering value for each mode. The solid lines indicate the case of a point absorber at the center of the fundamental mode, absorbing 10mW of incident power (although the normalization renders the absolute absorbed power irrelevant). 
The dashed lines show scattering from uniform absorption with an incident beam size of \SI{53}{\milli\meter}. 
Note that both cases were chosen to be axially symmetric about the center of the optic resulting in zero coupling to odd order modes, hence only even order modes are shown.

The time scale to achieve 70\% of the maximum scattering is strongly dependent on the mode order and only loosely dependent on the nature of absorption (uniform vs point absorber). 
This is understood given that (a) the time scale for thermal diffusion is a function of spatial scale and (b) the characteristic spatial scale of the nodes of  HOMs are a function of mode order (with higher order modes having smaller characteristic spatial scales). 
Hence shorter time constants function as an indicator for stronger prevalence of higher order modes.

Furthermore, the lower panel of Figure \ref{fig:scatter_to_HOM_v_t} shows the scattered into HOM from the fundamental mode in a steady-state single-bounce  (normalized to the power scattered into the second order mode). 
Scattering from a \SI{10}{\milli\watt} point absorber is indicated by the solid line and scattering from \SI{10}{\milli\watt} of  uniform absorption is indicated by the dashed line. 
As might be expected, the scattering from a point absorber contains substantially more HOM content (i.e. high spatial frequency content) than uniform absorption, with approximately $50$ times more power scattered into the 8th order mode in the point absorber case. 
Thus the stronger prevalence of higher order modes for a point absorber should yield shorter time constants for observed optical effects (than for uniform absorption). 
In an optical cavity, the exact scaling of time constants will be dependent on the resonant enhancement and suppression of higher order modes. This is explored in the next appendix.

%% file: IFOproperties.tex
\label{sec:IFOproperties}





In this section, we explore the interaction of power-dependent losses on the stored power of a resonant cavity. Although we describe the several power-dependent loss mechanisms that exist, the majority of this section focuses on resonant enhancement and suppression of losses from higher order modes which is the dominant loss mechanism.
Other adverse effects from power dependent losses are discussed in the following section, focusing on  intensity noise coupling. 



\begin{figure}[h]
  \centering
  \includegraphics[width=11cm]{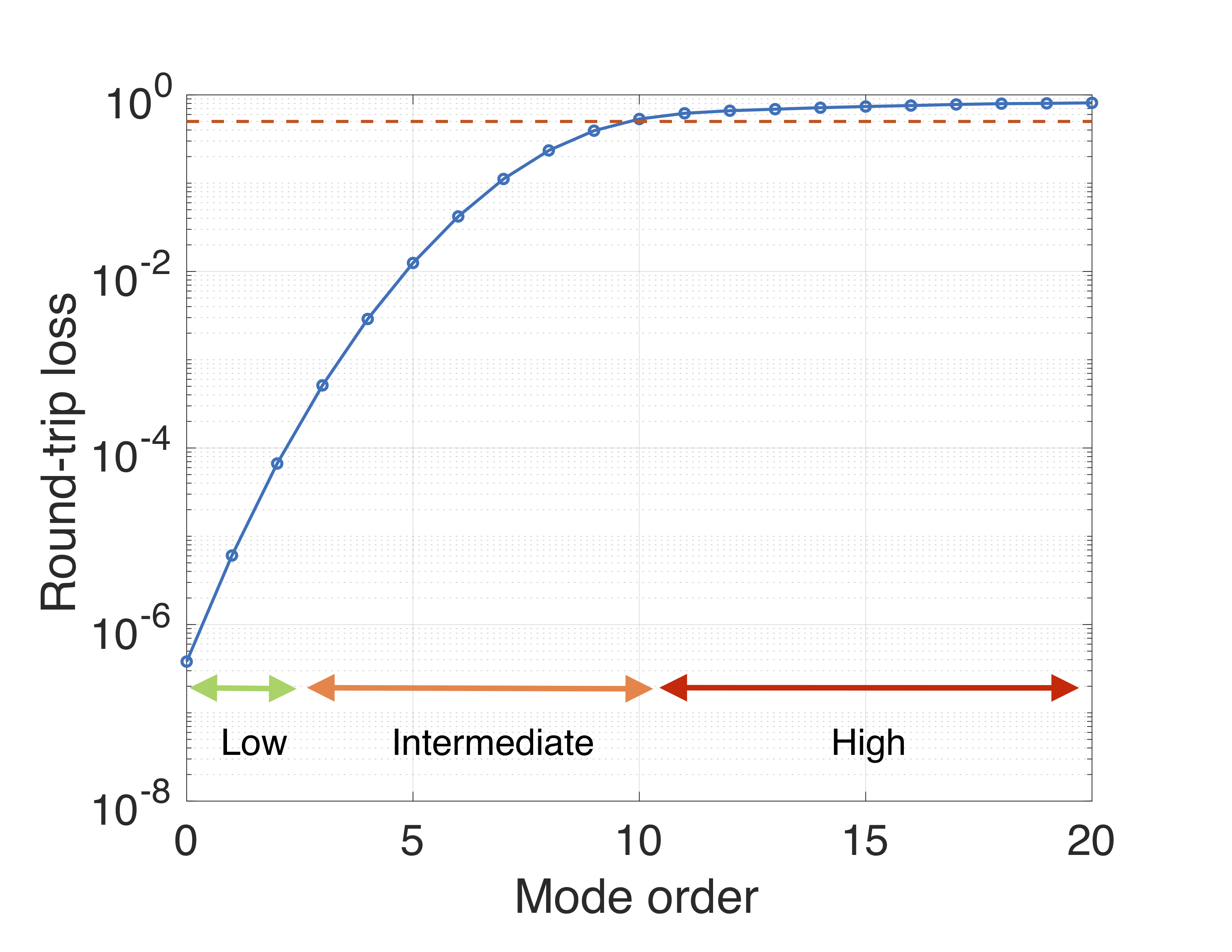}
  \caption{Round-trip loss, $\mathrm{L}_{RT_{m+n}}$, vs mode-order for a single round trip of higher order modes in the aLIGO arm cavities. The labels for "low", "intermediate" and "high" spatial frequencies are loosely defined and intended only as a guide to help associate different mode orders with different loss mechanisms.}
  \label{fig:RTLoss}
\end{figure}

\subsection{Loss mechanisms}


We characterize {\it power-dependent} losses from point absorbers into the following categories.

\begin{itemize}
    \item {\bf Wide-angle scattering to the beam tube}. These losses include all power lost on a single bounce of the carrier field from the surface of an optic that does not make it to the test mass at the other end of the arm.  For the context of this discussion, we define these as scatter into higher order modes where the round-trip clipping loss of that mode is greater than 50\%. This is illustrated in Figure \ref{fig:RTLoss}, and with the somewhat arbitrary delineation at mode order greater than 9 for aLIGO. Simulations show that the total power lost to these modes is minimal and thus we do not consider these in detail here.
    
    \item {\bf Resonant enhancement of losses from intermediate higher order modes}. These losses include scattering and resonant loss of power into higher order modes. These higher order modes experience more clipping at the edges of the test masses and hence more loss. In practice, these are due to low to medium spatial frequencies. This constitutes the main source of arm loss and is discussed in more detail in the following subsections.
    
    \item {\bf Mode-matching losses}.
        \begin{itemize}

    \item {\bf Differential (contrast defect) losses}. Differential mode changes between the two arms will increase loss to the anti-symmetric port of the interferometer.
    \item {\bf Common mode losses}. Common mode changes in the arms will reduce the overlap between the input laser mode and the resonant mode in the power-recycling cavity. 
    \end{itemize}
\end{itemize}

The last two items, contrast defect losses and reflection losses are, essentially, recycling cavity losses, $\mathcal{L}_{P}$ in Equation \ref{eqn:GP_gen}. 
When we model the aLIGO Fabry-Perot cavities with point absorbers, the mode-matching loss between the recycling cavities and the arms,  $\mathcal{L}_{P}$, are \DataVal{MM2ArmLossRatio}{} larger than arm cavity losses,  $\mathcal{L}_{A}$, in absolute terms. 
However, as indicated by Equation \ref{eqn:GP_gen}, the effect of arm losses on power-recycling are enhanced by the arm-cavity optical gain (a factor of \DataVal{ArmCavityFinesse}{}). 
Hence, the contribution of mode-mismatch between the two arms or arms and the power-recycling cavity is \DataVal{MM2ArmLossRatioWFinesse}{} smaller (i.e. are negligible or at most a perturbation) relative to the round-trip losses in the arms.

Thus, we shall discuss resonant enhancement and suppression of losses from intermediate higher order modes in more detail.



%% file: HOM_resonance_discussion.tex
\subsection{Scattering into non-resonant and partially resonant higher order modes}



The resonant enhancement and suppression of losses from the fundamental optical mode due to intra-cavity scattering  was previously solved by Vajente \cite{Vajente:14}. In that context, the solution was determined for small static deformations of mirror surfaces. We shall extend that analysis allowing for power-dependent surface deformations from point absorbers and for large surface aberrations present in real mirrors.


Equation (9)  from Vajente calculates the fundamental mode amplitude inside an optical  cavity that includes resonant scattering to a HOM. From that equation, the manuscript then determines the additional  loss term from the resonant scattering to a HOM in Equation (11) in that manuscript. 
Unfortunately, that manuscript contains a minor typographical error in the printed version of Equation (11) which does not affect its conclusions. We produce the correct derivation 
 below, split between Equations \ref{eqn:Gab_A} and \ref{eqn:Gab_B} and substituting some symbols for our own. The power loss from the fundamental mode to the $mn$-th higher order mode, $\mathcal{L}_{mn}$, is given by:


\begin{equation}
    \mathcal{L}_{mn} = a_{00|mn}^2\, g_{mn} \label{eqn:Gab_A}
\end{equation}

\noindent where $g_{mn}$ is the resonant enhancement/suppression factor of the $mn$-th higher order mode:

\begin{equation}
    g_{mn} = \frac{1 - r_{1}^{\prime \,2} \,r_{2}^{\prime \,2}}{1 + r_{1}^{\prime \,2}\,r_{2}^{\prime \,2} } \, \frac{1}{1 - \frac{2\, r_{1}^{\prime} \, r_{2}^{\prime} }{1 + r_{1}^{\prime\,2} \, r_{2}^{\prime\,2}} \,\cos \left[ \Phi_{mn}\right]} \label{eqn:Gab_B}
\end{equation}


\noindent where $r_{1^{\prime}}$ and $r_{2^{\prime}}$ are the modified amplitude reflectivities, attenuated by for mode-dependent clipping losses, of the ITM and ETM, respectively. 

\begin{equation}
    r_i^{\prime} = r_i\, \sqrt{ \frac{\int_0^{\Omega} \, r \, \left|E_{i,mn}\right|^2 \mathrm{d}r}{\int_0^{\infty} \, r \, \left|E_{i, mn}\right|^2 \mathrm{d}r} } \label{eqn:clipping_losses_HOM}
\end{equation}


\noindent where $\Omega$ is the radius of the mirror and $E_{i,mn}$ is the $mn$-th higher order mode propagated to the $i$-th mirror.



In the denominator of Equation \ref{eqn:Gab_B}, the argument of the cosine function, $\Phi_{mn}$, is an expression of the additional round-trip phase that higher order modes accumulate {\it relative to the fundamental mode}. For an ideal cavity with (a) infinite and (b) purely quadratic mirror surfaces, the nominal phase shift, $\Phi_{mn}^{\mathrm{nom}}$, is given by:

\begin{equation}
    \Phi_{mn}^{\mathrm{nom}} = (m+n)\, \phi_g \label{eqn:GPMpN}
\end{equation}

\noindent where $\phi_g$ is the round-trip Gouy phase shift \cite{Feng:01}, given by:

\begin{equation}
    \phi_g = 2\,\arccos\left[ \sqrt{\left(1-\frac{L}{\mathcal{R}_\mathrm{ITM}}\right) \, \left( 1-\frac{L}{\mathcal{R}_\mathrm{ETM}}\right)}\right], \label{eqn:nominal_gouy}
\end{equation}

where $\mathcal{R}_\mathrm{ITM}$ and $\mathcal{R}_\mathrm{ETM}$ are the ITM and ETM radii of curvature (ROC), respectively.





\begin{figure}[h]
  \centering
  \includegraphics[width=11cm]{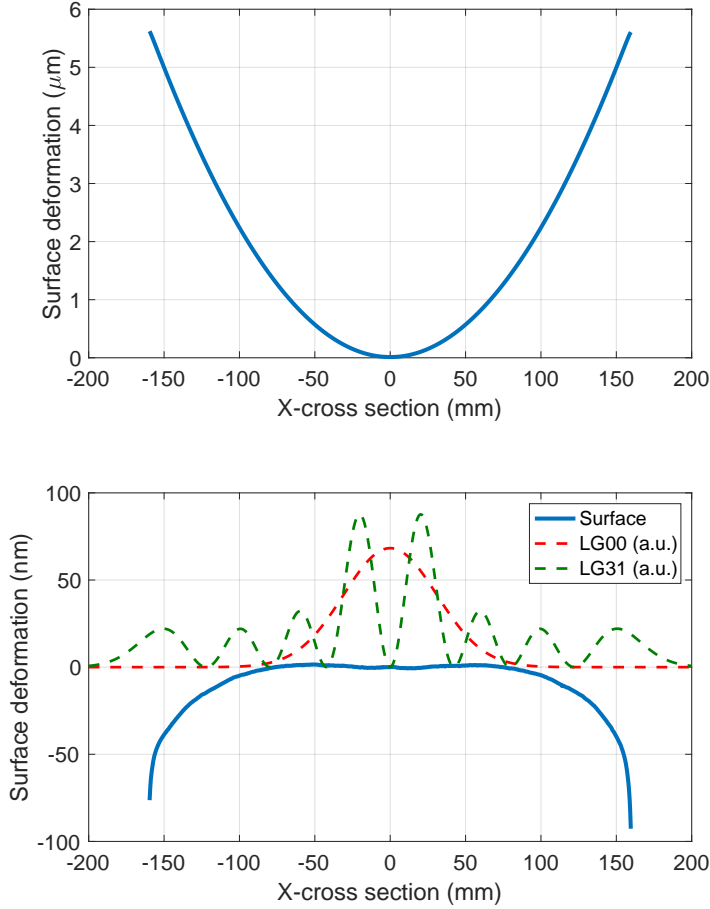}
  \caption{Optic cross section of ETMY at LHO. Upper panel shows a cross section of the surface of an ETM with a ROC of 2240m. The lower panel shows a cross section of residual ETM surface deformation after the spherical power term seen by the TEM00 mode is subtracted from the surface. For reference, intensity cross-sections of nominal LG00 modes and LG31 modes are shown. The zeroth order mode samples almost none of the residual distortion in the wings while the 7th order mode samples a large fraction of it.}
  \label{fig:ETM_cross_section}
\end{figure}


In a real interferometer such as Advanced LIGO, the mirror surfaces are (a) finite in diameter and (b) deviate from purely quadratic. This is illustrated in Figure \ref{fig:ETM_cross_section}
which shows a cross section of one of the LIGO ETMs in which the edges of the mirror deviate from quadratic by approximately \SI{50}{\nano\meter}. 
Higher order spatial modes with large spatial field extent, such as the Laguerre-Gauss three-one mode (LG31),  sample a larger fraction of this edge effect, relative to the fundamental mode, and accumulate an additional phase offset beyond the Gouy-phase expressed in Equation \ref{eqn:GPMpN}. 
This mirror phase offset/correction, $\Delta \phi_{mn}$, is a function of the mirror surface error, 
$S_{TM}$, and the spatial distribution of the HOM. Derived from the scattering of a HOM into itself:

\begin{equation}
    \left<\mathrm{E}_{mn}| \mathrm{e}^{\mathrm{i}\,2 \,k\,S_{TM}}|\mathrm{E}_{mn} \right> \approx  1 + 2\,\mathrm{i}\, \Delta \phi_{mn}
\end{equation}

\noindent where, $2\, \Delta \phi_{mn}$, the average phase accumulated by the mode is: 

\begin{equation}
     \Delta \phi_{mn} = \frac{2\,\pi}{\lambda}\, \frac{\iint \left|\mathrm{E}_{mn}\right|^2 \,S_{TM} \, \mathrm{d}x \,\mathrm{d}y }{\iint \left|\mathrm{E}_{mn}\right|^2 \, \mathrm{d}x \,\mathrm{d}y } \label{eqn:RT_phase_HOM_corr}
\end{equation}

\noindent  such that the round-trip phase accumulated by a higher order mode is given by:

\begin{equation}
    \Phi_{mn} = \Phi_{mn}^{\mathrm{nom}} + 2\, \Delta \phi_{mn} \label{eqn:total_RT_phase}
\end{equation}

\noindent where the leading factor of 2 is added to account for double-passing on reflection from the mirror. Note that we define the zero value of the residual surface error, $S_{TM}$, with such that  $\Delta \phi_{00} = 0$. The effect of the additional phase offset was studied by Bond et al \cite{Bond2011} in the context of a larger investigation exploring resonances higher order modes in optical cavities to reduce the effect of coating thermal noise in gravitational wave interferometers.

\begin{figure}[h]
  \centering
  \includegraphics[width=11cm]{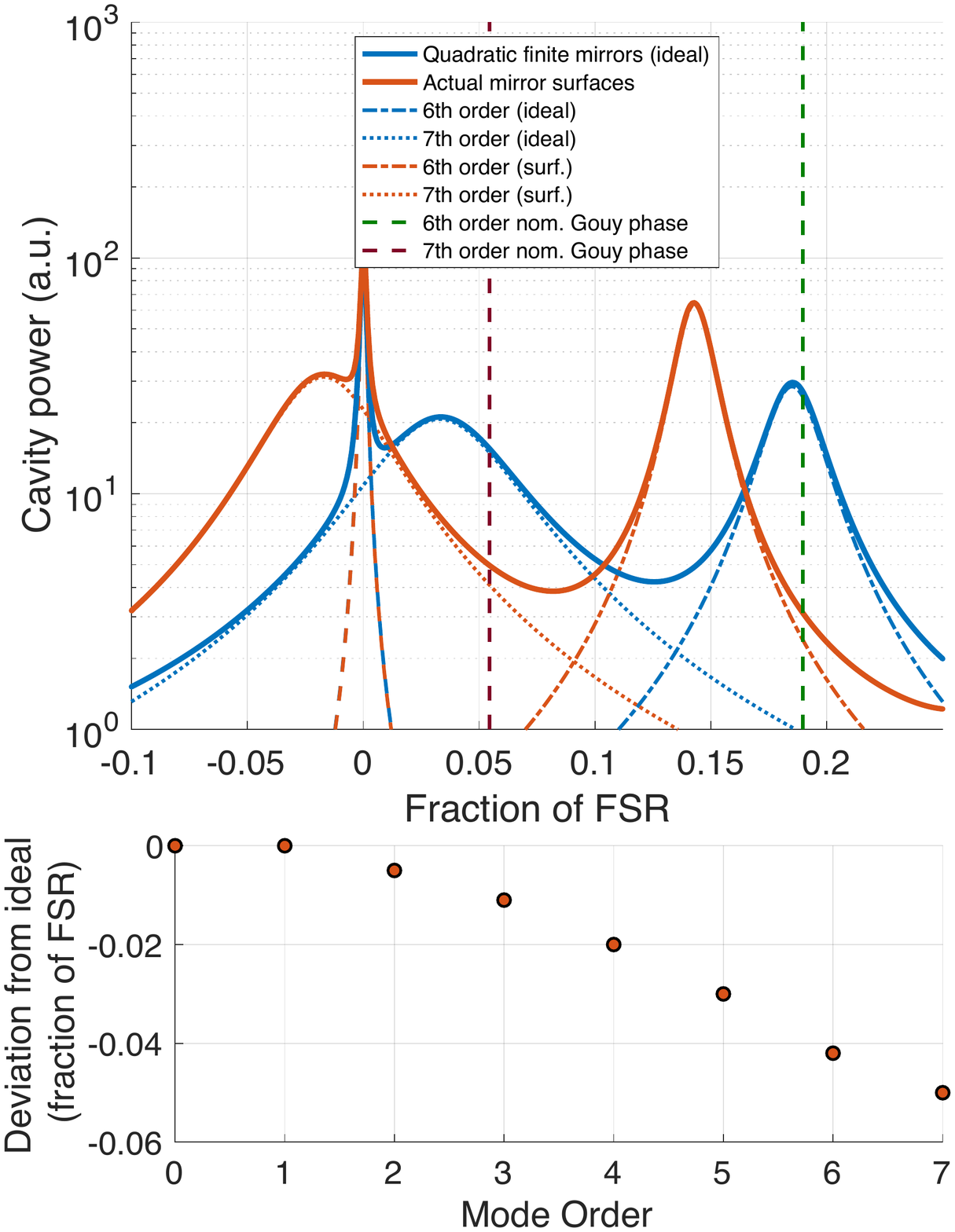}
  \input{ideal_v_real_caption}
  \label{fig:HOM_spacing}
\end{figure}

This residual surface error effect shifts the resonance location of modes from the nominal locations (calculated solely from the Gouy phase). This is illustrated in Figure \ref{fig:HOM_spacing}. 
The upper panel shows simulated cavity scans of a poorly matched beam into the LIGO-Hanford \SI{4}{\kilo\meter} Fabry-Perot X-arm cavity. 
The simulated beam is poorly matched in order to excite higher order modes in the scan. 
The blue and red curves show  results from purely quadratic mirrors and with included surface figure errors, respectively. 
The dot-dash and dotted lines show the 6th and 7th modes, respectively. 

Several features of this figure are notable. 
Firstly, the full-width half-maximum linewidths of the fitted 6th and 7th order modes are \DataVal{TEM60LineWidth}{} and \DataVal{TEM70LineWidth}{}  of an FSR, respectively. These are  considerably wider than the linewidth of the fundamental mode (\DataVal{TEM00LineWidth}{}  of an FSR), due to the increased round-trip loss, $\mathrm{L}_{RT_{m+n}}$, that HOMs experience from the finite mirror apertures (illustrated in Figure \ref{fig:RTLoss}).

Secondly, the actual resonance locations of the HOM have shifted relative to the case of ideal quadratic mirrors by amounts $\Delta\phi\left(E_{i}, S_\mathrm{TM}\right)$
In the case of the 7th order mode, the resonance has shifted past the fundamental mode resonance through to the other side. 
Note that in the ideal quadratic mirror case (blue), there is still an anomalous phase shift: the 7th order resonance has shifted approximately 2\% of an FSR relative to the resonance calculated from the nominal Gouy phase (dashed purple). This is addressed further when we discuss the limitations of the resonant analysis detail in Equations \ref{eqn:Gab_A} through \ref{eqn:total_RT_phase}.
The bottom panel shows the difference in the resonance locations of the higher order modes for the two cases. 

\begin{figure}[h]
 \centering
 \includegraphics[width=11cm]{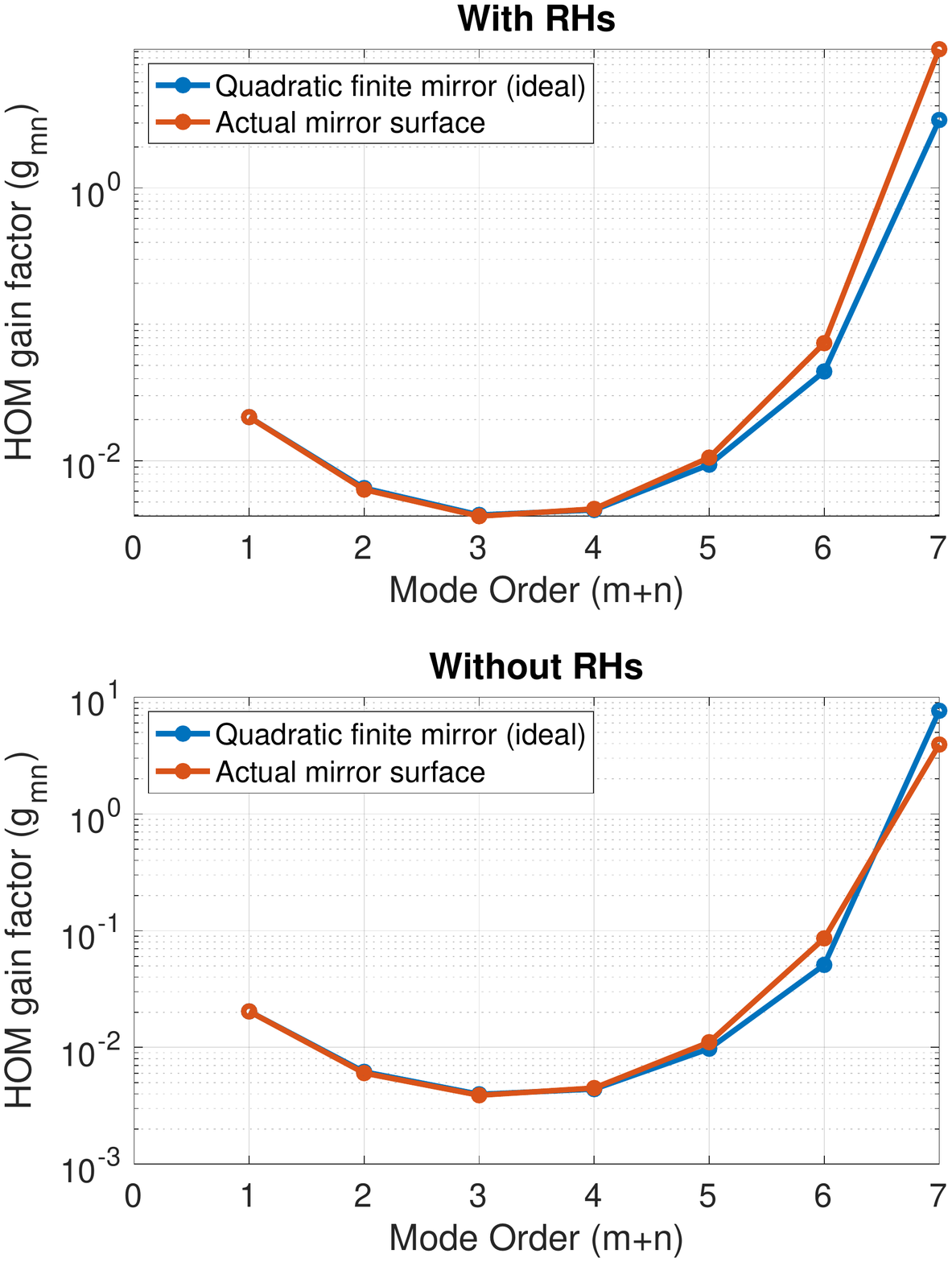}
 \caption{Higher order mode (HOM) gain factor for the LHOX aLIGO cavity modeled for quadratic finite mirrors (i.e. both ITM and ETM are purely quadratic) and including the actual mirror surfaces 
 (both ITM and ETM surface polishing maps are included). Top plot: with the ring heaters at O3 settings, mirror radii of curvature, ITM = \SI{1939.17}{\meter}, ETM = \SI{2239.65}{\meter}.  Bottom plot: ring heaters off and mirrors at manufactured ROC, ITM = \SI{1940.3}{\meter}, ETM = \SI{2244.2}{\meter}.}
 \label{fig:HOM_gain}
\end{figure}

The resonant enhancement/suppression (HOM gain factor $g_{mn}$) is shown for the aLIGO arm cavity in Figure \ref{fig:HOM_gain}. It is notable that simply adding  the real mirror surfaces error, a cross-section of which is illustrated in Figure \ref{fig:ETM_cross_section}, amplifies the loss by a factor of 3 for the 7-th order mode. However, this effect is also dependent on the ring heater settings (more precisely, the mirror ROC), as illustrated by the bottom panel of Figure \ref{fig:HOM_gain} in which the addition of the actual mirror surface reduces the overall gain of the 7-th order mode when the ring heaters are not employed. 

Thus, having determined the total round-trip loss from the Vajente formalism expressed in Equations \ref{eqn:Gab_A} and \ref{eqn:Gab_B} and accounting HOM round-trip phase perturbations due to residual   surface   errors in the mirrors, expressed in Equation \ref{eqn:RT_phase_HOM_corr}, one can approximate the impact on the power recycling gain. 

One final note on the applicability of this resonant analysis. There are several implicit assumptions in the Vajente formalism that limit its predictive ability. Specifically, we assume:
\begin{itemize}
    \item power is scattered between the fundamental and individual higher order modes only, and not between different higher order modes. This assumption becomes less valid for higher order modes with significant clipping at the edge of the mirror. 
    This manifests in Figure \ref{fig:HOM_spacing}, the location of the 7th order mode for the simulation of ideal quadratic mirrors (blue)  shows an anomalous phase shift to the left relative to the resonant location of the 7th order mode calculated from the Gouy phase (purple). This is independent of the surface figure of the mirrors. This anomalous phase shift is due to power from the 7th order mode scattering into HOM (due to the aperture of the mirror), propagating through a round-trip of the cavity, where it accumulates  additional Gouy phase, and then scattering back into the 7th order mode after one (or more) round trips. The resonant analysis presented in this paper does not account for this mode-mixing: that is, in a full mode-mixing analysis, there will be an additional phase term in Equation \ref{eqn:total_RT_phase} that accounts for this.
    \item that the loss is the same for modes of the same order - as expressed in Figure \ref{fig:RTLoss}. This is appropriate on average but the exact losses for different HOM will depend on the overlap between the mirror shape and the individual mode shape. 
    \item that the HOM phase shift is the same for all HOM of the same order. In reality, there will be small variations between modes. This will result in mode-splitting, but the phase variations between different modes will be small relative to the broad linewidths. This will manifest as a broadening of a resonant peak of modes of a specific order.
\end{itemize}

Therefore, the formalism presented here is valid for rapid estimation of the effects on power-recycling gain. The {\it relative} shift in resonance location due to the surface figure can be computed using this formalism. 
However, the {\it absolute} resonance locations of higher order modes should be treated carefully as mode-mixing induced phase shifts are not accounted for in this formalism. Specifically, the mode-mixing induced phase shift would need to be added to Equation \ref{eqn:total_RT_phase} for absolute calculations of gain. For more precise calculations, more comprehensive numerical/analytic simulations are recommended. Suggested examples are  Finesse \cite{FinesseRef} code, a multi-modal analysis of optical cavities and SIS (Static Interferometer Simulation), an FFT-based numerical analysis of optical cavities \cite{SIS-T070039}. 

It is worth emphasizing that the results in Figure \ref{fig:HOM_gain} underscore that precise ROC and mirror surface errors must be considered when determining the resonant gain of HOM in the cavity. Indeed, numerical calculations by SIS show that the O3 ETM surface maps suppress the gain of the 7th mode relative to the spherical shape, and will suppress the round trip loss. 
This is used to design the optimal surface shape of the O4 ETM maps to mitigate the effect of point absorbers.


%% file: ideal_v_real_caption.tex
  \caption{Top: modeled cavity scan of the LHOX aLIGO arm cavity including ring heater effects (ITMROC = 1939.17m, ETMROC = 2239.65m), with ideal quadratic finite diameter mirrors (blue) and with actual mirror surface deformation included (red).
 A highly mismatched beam (composed of many cavity HOM) was injected into the model to better highlight the location of the different higher order modes: 6th order (dot-dash), 7th order (dotted). 
 Bottom: the deviation of the resonance location of the higher order modes [expressed as a fraction of a free-spectral range (FSR)]. 
}

%% file: PRG_discussion.tex
\subsection{Power recycling gain and generalized arm loss.}
\label{sec:prg_generalized_discussion}

Thus, to summarize, the full arm loss, $\mathcal{L}_A$, is dependent on the elements described in Appendices \ref{sec:Optics}, \ref{sec:single-bounce}, \ref{sec:IFOproperties}. 
Bringing all these concepts together allows us to better estimate the power recycling gain, our proxy for overall interferometer performance at high frequencies, in the presence of losses due to point absorbers. 

As expressed in Equation \ref{eqn:Gab_A}, the overall loss is a function of the single-bounce amplitude scattering into HOM and the resonant enhancement or suppression of those modes, the elements of which are summarized below:

\begin{enumerate}
    \item Single-bounce amplitude scattering depends upon:
    
    \begin{enumerate}
        \item Total absorbed power: $P_\mathrm{abs}$, (Equation \ref{eqn:outside_abs}). For a LIGO arm cavity this is, in turn, a product of the arm power, $P_A$, the  incident laser beam size, $\mathrm{w}$, the location of the point absorber, $\mathbf{r_c}$, and the equivalent ``100\% absorbing area'' of the point absorber, (Equation \ref{eqn:power_abs}).
        \item Material properties: thermal conductivity, $\kappa$, and the coefficient of thermal expansion, $\alpha$, (Equation \ref{eqn:outside_abs}).
        \item Location of the point absorber: $\mathbf{r_c}$, from the context of its position relative to HOM field distribution, (Equation \ref{eqn:scatt_posn_dependence}).
    \end{enumerate}
    
    \item Resonant enhancement or suppression depends upon:
    
    \begin{enumerate}
        \item Clipping losses in HOM, $r_{i}^{\prime}$, a function of mode size and mirror size (Equation \ref{eqn:clipping_losses_HOM}).
        \item Nominal round-trip Gouy phase: $\phi_g$, a function of cavity geometry (Equation \ref{eqn:nominal_gouy})
        \item Additional round-trip phase perturbations from mirror surface errors: $\Delta \phi_{mn}$ (Equation \ref{eqn:RT_phase_HOM_corr}).
    \end{enumerate}
\end{enumerate}

Which, when combined,  averaged across the two interferometer arms and accounting for nominal power loss, the total arm loss can be expressed as:

 \begin{eqnarray}
      \mathcal{L}_\mathrm{A} & = & \mathcal{L}_\mathrm{nom} + \frac{1}{2} \sum_{m,n}  \mathcal{L}_{mn,X}  + \mathcal{L}_{mn,Y} \label{eqn:loss_avg1_app}\\
       & = & \mathcal{L}_\mathrm{nom}+ b\, \left(\frac{P_A}{100\mathrm{kW}}\right)^2 \label{eqn:loss_avg2_app}
       \end{eqnarray}

\noindent where $\mathcal{L}_\mathrm{nom}$ is the nominal arm loss and $b$ is the coefficient for power-dependent loss  encompassing the elements described above.
As shown in Equation \ref{eqn:a00mm_w_fr}, all the amplitude scattering coefficients contain a common factor of absorbed power, $P_\mathrm{abs}$, which is, in turn dependent on arm power, $P_A$. Thus, we can finally write the power-recycling gain as a function of arm power:

\begin{equation}
       G_P =  \left( \frac{t_p}{1-r_p\,  \left(1 - G_A\,\left[\mathcal{L}_\mathrm{nom} + b\, \left(\frac{P_A}{100\mathrm{kW}}\right)^2\right]\right)} \right)^2 \label{eqn:PRG_Psq}
\end{equation}

It is the $b$ term that must be minimized to maximize the power-recycling gain.

One final comment regarding extensions to this analysis. This discussion focused on a single absorber on a single optic. It is easily extended to multiple point absorbers since, as shown in Equation \ref{eqn:linear_amp_scatt}, amplitude scattering to HOM is linear with respect to surface deformations that are small compared to the wavelength of light. 
Thus the effect of multiple point absorbers can be determined by taking a linear sum of amplitude scattering coefficients from individual points.

%% file: ITMeffects.tex
\label{sec:ITMeffects}

Power dependent losses from surface deformation are the main observed effect of thermal effects point absorbers.  As an example of secondary adverse effects due to point absorbers (and ones which originate from substrate thermal lenses), this appendix describes why we expect increased coupling of intensity noise to the gravitational wave channel via the feedback control system.

\subsection{Substrate thermal lens effect of carrier and sidebands}




Due to the complexity of the Advanced LIGO length sensing and control system \cite{aLIGO_Martynov_2016, aLIGO_ASC_Barsotti_2010, Staley:15, Mullavey:12}, we limit discussion of the interaction of that system with point absorbers to the  case study of increased relative intensity noise coupling.


As described in Appendix \ref{sec:aLIGOOperation_2}, a feedback control system, using frontal modulation RF sidebands and the Pound-Drever-Hall (PDH) technique, keeps the interferometer  operating at the correct working point to ensures high sensitivity to gravitational waves. 

In the ideal scenario, only the carrier fundamental mode, $\modesSym{}^{C}_{00}$, and the sideband fundamental mode, $\modesSym{}^{SB}_{00}$, beat against each other on detection photodiodes to create PDH control signals. The diodes sees one large PDH signal  centered  around  the correct locking  point. 

Any optical path distortion (OPD) in the input test mass substrate affects both the carrier and sideband fields in the single propagation through the optics. 
If we assume that the field at the AR face of the ITM has a TEM00 Gaussian mode distribution, matched to the cavity resonant mode, then at the lowest order the effect of the ITM distortion is to couple a fraction, $\epsilon$, of this fundamental mode into a combination of higher order modes:
\begin{equation}
    \modesSym{}_{ITM} = \sqrt{1 - |\epsilon|^2} \modesSym{}_{00} + \epsilon\, \modesSym{}_{HOM}
\end{equation}
This equation is valid for carrier and sideband fields equally. When considering the reflection off the Fabry-Perot arm cavity, we have to distinguish between the carrier (TEM00 mode is resonant and HOM are antiresonant) and the sidebands (both TEM00 and HOM modes are antiresonant). 
A field resonant in the arm cavity is reflected with a  phase shift of $\pi$ radians, while a field antiresonant in the arm cavity is reflected with zero phase shift. Therefore the fields coming back from the cavities, but before passing through the ITM substrate, are
\begin{eqnarray}
    \modesSym{}_{R}^{C} &=& -\sqrt{1 - |\epsilon|^2} \,\modesSym{}_{00}^C + \epsilon\, \modesSym{}_{HOM}^C \\
    \modesSym{}_{R}^{SB} &=& \sqrt{1 - |\epsilon|^2} \,\modesSym{}_{00}^{SB} + \epsilon\, \modesSym{}_{HOM}^{SB} 
\end{eqnarray}
The transmission through the ITM distorted substrate couples the same amount, $\epsilon$, of TEM00 into the same high order mode as for the first pass. But since the carrier reflected back has the opposite sign, at first order in the OPD the carrier HOM field cancel outs. This is not the case for the sideband fields.

Therefore the carrier is largely immune to the effect of the optical path distortion caused by point absorbers 
while the sidebands are fully affected.

\subsection{Impact on control system and elevated intensity noise coupling}



Now, consider the input intensity noise carried on the laser fields. Only the carrier fundamental mode, $\modesSym{}^{C}_{00}$,  is resonant in the arm cavity and, as such, is the only field to experience the intensity noise filtering effect of that cavity (a pole around \SI{0.6}{\hertz} \cite{Martynov_noise:15}).
The remaining fields, $\modesSym{}^{C}_{\mathrm{HOM}}$, $\modesSym{}^{SB}_{00}$ and $\modesSym{}^{SB}_{\mathrm{HOM}}$, experience no optical suppression of intensity noise. 

The carrier HOM and fundamental fields do not directly beat against each other and, hence, do not contribute to RIN coupling. However, as shown in Figure \ref{fig:IFO_schematic}, the output of the interferometer passes through an output mode cleaner (OMC). 
Any mode-mismatch between the interferometer beam and the OMC \cite{AWC2020} will allow some carrier HOM to mix with the fundamental and present a path for intensity noise to couple to the detection photodiodes. This is one possible way an OPD can increase the coupling of intensity noise (active wavefront control are being commissioned to minimize this mode-mismatch \cite{Cao:20}).

Additionally, indirect coupling of laser intensity noise is created due to the interaction of carrier and sideband HOM with the feedback control systems. These two fields will beat against each other creating {\it additional} RF PDH-like signal on the control photodiodes.  
Due to Gouy phase differences, these are offset from nominal, or zero, locking point. 

A offset in the short Michelson control signal means that the IFO is operating such that the two short Michelson arms are not "equal" to get perfect destructive interference.
However, the differential arm (DARM) loop is locked  to keep the output power to be a constant value \cite{Fricke_2012}.
When coupled with the  Michelson offset, the DARM loop is driven offset from the "proper" working point, so to produce another static carrier field imbalance that cancel out the field created by the MICH offset.
Although the static power cancels out, the RIN on the two carrier fields do not cancel out, since one is noisy (coming from PRC/MICH) and the other is clean (coming from arms). 
The end result is an increase in intensity noise coupling through the dark port, still mediated by TEM00 modes, and so not "fixed" by the OMC.


%% file: parameters.tex
\begin{table}[htp]
  \centering
  \begin{tabular}{l | c | c} 
{\bf Parameter} & {\bf Symbol} & {\bf Value} \\ 
 \hline\hline
 ETM transmission (nom.) & $T_e$ ($t_e^2$) & 5 ppm  \\ 
 ITM transmission (nom.) & $T_i$ ($t_i^2$) & 1.4\%  \\ 
 PRM transmission (nom.) & $T_p$ ($t_p^2$) & 3.0\%  \\ 
 \hline
 Arm optical gain (nom.) & $G_A$ & $\approx$ 280 \\
 Arm optical gain (LHO) & $G^{H}_A$ & 268 \cite{O3CommissioningPRD}\\
 Arm optical gain (LLO) & $G^{L}_A$ & 265 \cite{O3CommissioningPRD} \\
 \hline
 ITM TEM00 beam rad. & $\mathrm{w}_i$ & \SI{53}{\milli\meter}  \\ 
 ETM TEM00 beam rad. & $\mathrm{w}_e$ & \SI{62}{\milli\meter}  \\ 
 \hline
 Test mass (TM) radius  & $\Omega$ & \SI{170}{\milli\meter}  \\ 
 TM thickness & $h$ & \SI{200}{\milli\meter}  \\ 
 Thermal expansion coeff.  & $\alpha$ & \SI{0.55E-6}{\per\kelvin}  \\ 
 Thermal conductivity & $\kappa$ & \SI{1.38}{\watt\per\meter\per\kelvin}  \\ 
 TM specific heat capacity & $c$ & \SI{703}{\joule\per\kilogram\per\kelvin} \\ 
 TM density & $\rho$ & \SI{2203}{\kilogram\per\meter^3}  \\ 
 \hline
 Laser wavelength & $\lambda$ & \SI{1064}{\nano\meter} \\
 Arm length & $L$ & \SI{3995}{\meter} \\
 ITM ROC (nom.) & $\mathcal{R}_{ITM}$ & \SI{1939}{\meter} \\
 ETM ROC (nom.) & $\mathcal{R}_{ETM}$ & \SI{2245}{\meter} \\ \hline
 ITM ROC (LHOX wo. RH) & $\mathcal{R}^{H*}_{ITM}$ & \SI{1940.3}{\meter} \\
 ETM ROC (LHOX wo. RH) & $\mathcal{R}^{H*}_{ETM}$ & \SI{2244.2}{\meter} \\ 
 ITM ROC (LHOX w. RH) & $\mathcal{R}^{H}_{ITM}$ & \SI{1939.17}{\meter} \\
 ETM ROC (LHOX w. RH) & $\mathcal{R}^{H}_{ETM}$ & \SI{2239.65}{\meter} \\ 
  \hline
  \end{tabular}
  \caption{Relevant parameters for aLIGO. The specific ROC for LHOX account include the change due to the application of the TCS ring heater (RH). }
  \label{tab:values}
\end{table}



%% file: PTAbs.bbl
\begin{thebibliography}{10}
\expandafter\ifx\csname urlstyle\endcsname\relax
  \providecommand{\doi}[1]{doi:\discretionary{}{}{}#1}\else
  \providecommand{\doi}{doi:\discretionary{}{}{}\begingroup
  \urlstyle{rm}\Url}\fi

\bibitem{O3CommissioningPRD}
A.~Buikema, et~al.
\newblock Sensitivity and performance of the advanced ligo detectors in the
  third observing run.
\newblock \emph{Phys. Rev. D}, 102:062003, Sep 2020.
\newblock \doi{10.1103/PhysRevD.102.062003}.

\bibitem{GW150914etal}
B.~P. Abbott et~al.
\newblock Observation of gravitational waves from a binary black hole merger.
\newblock \emph{Phys. Rev. Lett.}, 116:061102, Feb 2016.
\newblock \doi{10.1103/PhysRevLett.116.061102}.

\bibitem{O1O2Catalog}
B.~P. Abbott, et~al.
\newblock Gwtc-1: A gravitational-wave transient catalog of compact binary
  mergers observed by ligo and virgo during the first and second observing
  runs.
\newblock \emph{Phys. Rev. X}, 9:031040, Sep 2019.
\newblock \doi{10.1103/PhysRevX.9.031040}.

\bibitem{Note1}
Https://gracedb.ligo.org/superevents/public/O3/.

\bibitem{aLIGO_Overview}
{The LIGO Scientific Collaboration}.
\newblock Advanced {LIGO}.
\newblock \emph{Classical and Quantum Gravity}, 32(7):074001, 2015.

\bibitem{Corbitt_2004}
T.~Corbitt et~al.
\newblock Review: Quantum noise in gravitational-wave interferometers.
\newblock \emph{Journal of Optics B: Quantum and Semiclassical Optics},
  6(8):S675--S683, jul 2004.
\newblock \doi{10.1088/1464-4266/6/8/008}.

\bibitem{PhysRevA.44.7022}
W.~Winkler, et~al.
\newblock Heating by optical absorption and the performance of interferometric
  gravitational-wave detectors.
\newblock \emph{Phys. Rev. A}, 44:7022--7036, Dec 1991.
\newblock \doi{10.1103/PhysRevA.44.7022}.

\bibitem{HelloVinet_TE:90}
{Hello, Patrice} et~al.
\newblock Analytical models of transient thermoelastic deformations of mirrors
  heated by high power cw laser beams.
\newblock \emph{J. Phys. France}, 51(20):2243--2261, 1990.
\newblock \doi{10.1051/jphys:0199000510200224300}.

\bibitem{HelloVinet:90_TL}
{Hello, Patrice} et~al.
\newblock Analytical models of thermal aberrations in massive mirrors heated by
  high power laser beams.
\newblock \emph{J. Phys. France}, 51(12):1267--1282, 1990.
\newblock \doi{10.1051/jphys:0199000510120126700}.

\bibitem{Brooks:16}
A.~F. Brooks, et~al.
\newblock Overview of {Advanced LIGO} adaptive optics.
\newblock \emph{Appl. Opt.}, 55(29):8256--8265, Oct 2016.
\newblock \doi{10.1364/AO.55.008256}.

\bibitem{Brooks:09}
A.~F. Brooks, et~al.
\newblock Direct measurement of absorption-induced wavefront distortion in high
  optical power systems.
\newblock \emph{Appl. Opt.}, 48(2):355--364, Jan 2009.
\newblock \doi{10.1364/AO.48.000355}.

\bibitem{Note2}
{https://dcc.ligo.org/LIGO-G1900693}.

\bibitem{Vajente:14}
G.~Vajente.
\newblock In situ correction of mirror surface to reduce round-trip losses in
  {Fabry-Perot} cavities.
\newblock \emph{Appl. Opt.}, 53(7):1459--1465, Mar 2014.
\newblock \doi{10.1364/AO.53.001459}.

\bibitem{Note3}
D. Martynov, {LLO aLOG 43121, February 2019},
  {https://alog.ligo-la.caltech.edu/aLOG/ }.

\bibitem{Note4}
{https://labcit.ligo.caltech.edu/~hiro/SIS/}.

\bibitem{Vajente_2013}
G.~Vajente.
\newblock Fast modal simulation of paraxial optical systems: the {MIST} open
  source toolbox.
\newblock \emph{Classical and Quantum Gravity}, 30(7):075014, mar 2013.
\newblock \doi{10.1088/0264-9381/30/7/075014}.

\bibitem{CHRAAC:2013}
R.~A. Day, et~al.
\newblock Reduction of higher order mode generation in large scale
  gravitational wave interferometers by central heating residual aberration
  correction.
\newblock \emph{Phys. Rev. D}, 87:082003, Apr 2013.
\newblock \doi{10.1103/PhysRevD.87.082003}.

\bibitem{Voyager_2020}
R.~X. Adhikari, et~al.
\newblock A cryogenic silicon interferometer for gravitational-wave detection.
\newblock \emph{Classical and Quantum Gravity}, 37(16):165003, jul 2020.
\newblock \doi{10.1088/1361-6382/ab9143}.

\bibitem{NEMO:2020}
K.~Ackley et~al.
\newblock {Neutron Star Extreme Matter Observatory}: A kilohertz-band
  gravitational-wave detector in the global network, arxiv: 2007.03128.
\newblock \emph{preprint}, 2020.

\bibitem{Hiro2020}
M.~V. Poplavskiy, et~al.
\newblock Diffraction losses of a fabry-perot cavity with nonidentical
  non-spherical mirrors.
\newblock \emph{preprint, https://arxiv.org/abs/2005.02033}, 2020.

\bibitem{Miao_HF:2018}
H.~Miao, et~al.
\newblock Towards the design of gravitational-wave detectors for probing
  neutron-star physics.
\newblock \emph{Phys. Rev. D}, 98:044044, Aug 2018.
\newblock \doi{10.1103/PhysRevD.98.044044}.

\bibitem{Martynov:2019_PRD}
D.~Martynov, et~al.
\newblock Exploring the sensitivity of gravitational wave detectors to neutron
  star physics.
\newblock \emph{Phys. Rev. D}, 99:102004, May 2019.
\newblock \doi{10.1103/PhysRevD.99.102004}.

\bibitem{Page_HF_GW:2020}
M.~A. Page, et~al.
\newblock Gravitational wave detectors with broadband high frequency
  sensitivity.
\newblock \emph{preprint}, 2020.

\bibitem{ET2010}
M.~Punturo, et~al.
\newblock The third generation of gravitational wave observatories and their
  science reach.
\newblock \emph{Classical and Quantum Gravity}, 27(8):084007, apr 2010.
\newblock \doi{10.1088/0264-9381/27/8/084007}.

\bibitem{Reitze2019Cosmic}
D.~Reitze, et~al.
\newblock {Cosmic Explorer}: The {U.S}. contribution to gravitational-wave
  astronomy beyond {LIGO}.
\newblock \emph{Bulletin of the AAS}, 51(7), 9 2019.
\newblock Https://baas.aas.org/pub/2020n7i035.

\bibitem{aLIGO_ASC_Barsotti_2010}
L.~Barsotti, et~al.
\newblock Alignment sensing and control in advanced {LIGO}.
\newblock \emph{Classical and Quantum Gravity}, 27(8):084026, apr 2010.
\newblock \doi{10.1088/0264-9381/27/8/084026}.

\bibitem{aLIGO_Staley_2014}
A.~Staley, et~al.
\newblock Achieving resonance in the advanced {LIGO} gravitational-wave
  interferometer.
\newblock \emph{Classical and Quantum Gravity}, 31(24):245010, nov 2014.
\newblock \doi{10.1088/0264-9381/31/24/245010}.

\bibitem{aLIGO_Martynov_2016}
D.~V. Martynov, et~al.
\newblock Sensitivity of the advanced ligo detectors at the beginning of
  gravitational wave astronomy.
\newblock \emph{Phys. Rev. D}, 93:112004, Jun 2016.
\newblock \doi{10.1103/PhysRevD.93.112004}.

\bibitem{aLIGO_LSC_2016}
B.~P. Abbott, et~al.
\newblock Gw150914: The advanced ligo detectors in the era of first
  discoveries.
\newblock \emph{Phys. Rev. Lett.}, 116:131103, Mar 2016.
\newblock \doi{10.1103/PhysRevLett.116.131103}.

\bibitem{aLIGO_O2_Driggers}
J.~C. Driggers, et~al.
\newblock Improving astrophysical parameter estimation via offline noise
  subtraction for advanced ligo.
\newblock \emph{Phys. Rev. D}, 99:042001, Feb 2019.
\newblock \doi{10.1103/PhysRevD.99.042001}.

\bibitem{Siegman:Lasers}
A.~Siegman.
\newblock \emph{Lasers}.
\newblock University Science Books, 1986.
\newblock ISBN 9780935702118.

\bibitem{Martynov_noise:15}
D.~V. Martynov, et~al.
\newblock Sensitivity of the advanced ligo detectors at the beginning of
  gravitational wave astronomy.
\newblock \emph{Phys. Rev. D}, 93:112004, Jun 2016.
\newblock \doi{10.1103/PhysRevD.93.112004}.

\bibitem{Hild_2009}
S.~Hild, et~al.
\newblock {DC}-readout of a signal-recycled gravitational wave detector.
\newblock \emph{Classical and Quantum Gravity}, 26(5):055012, feb 2009.
\newblock \doi{10.1088/0264-9381/26/5/055012}.

\bibitem{RSE:2003}
J.~E. Mason et~al.
\newblock Signal extraction and optical design for an advanced
  gravitational-wave interferometer.
\newblock \emph{Appl. Opt.}, 42(7):1269--1282, Mar 2003.
\newblock \doi{10.1364/AO.42.001269}.

\bibitem{AndoThesis}
M.~Ando.
\newblock \emph{Power Recycling for an Interferometric Gravitational Wave
  Detector}.
\newblock Ph.D. thesis, University of Tokyo, 1998.

\bibitem{PDH_1983}
R.~W.~P. Drever, et~al.
\newblock Laser phase and frequency stabilization using an optical resonator.
\newblock \emph{Applied Physics B}, 31(2):97--105, 1983.
\newblock \doi{10.1007/BF00702605}.

\bibitem{HallThesis_2017}
E.~Hall.
\newblock \emph{Long-baseline laser interferometry for the detection of binary
  black-hole mergers}.
\newblock Ph.D. thesis, {California Institute of Technology}, 2017.

\bibitem{MartynovThesis}
D.~Martynov.
\newblock \emph{Lock Acquisition and Sensitivity Analysis of Advanced LIGO
  Interferometers}.
\newblock Ph.D. thesis, {California Institute of Technology}, 2015.

\bibitem{Feng:01}
S.~Feng et~al.
\newblock Physical origin of the gouy phase shift.
\newblock \emph{Opt. Lett.}, 26(8):485--487, Apr 2001.
\newblock \doi{10.1364/OL.26.000485}.

\bibitem{Bond2011}
C.~Bond, et~al.
\newblock Higher order laguerre-gauss mode degeneracy in realistic, high
  finesse cavities.
\newblock \emph{Phys. Rev. D}, 84:102002, Nov 2011.
\newblock \doi{10.1103/PhysRevD.84.102002}.

\bibitem{FinesseRef}
A.~Freise, et~al.
\newblock Finesse, {Frequency domain INterferomEter Simulation SoftwarE}.
\newblock Technical report, \url{https://arxiv.org/abs/1306.2973}, 2013.

\bibitem{SIS-T070039}
H.~Yamamoto.
\newblock {SIS (Stationary Interferometer Simulation) manual}.
\newblock Technical report, LIGO Laboratory,
  \url{https://dcc.ligo.org/LIGO-T070039/public}, 2013.

\bibitem{Staley:15}
A.~Staley, et~al.
\newblock High precision optical cavity length and width measurements using
  double modulation.
\newblock \emph{Opt. Express}, 23(15):19417--19431, Jul 2015.
\newblock \doi{10.1364/OE.23.019417}.

\bibitem{Mullavey:12}
A.~J. Mullavey, et~al.
\newblock Arm-length stabilisation for interferometric gravitational-wave
  detectors using frequency-doubled auxiliary lasers.
\newblock \emph{Opt. Express}, 20(1):81--89, Jan 2012.
\newblock \doi{10.1364/OE.20.000081}.

\bibitem{AWC2020}
A.~Perreca, et~al.
\newblock Analysis and visualization of the output mode-matching requirements
  for squeezing in advanced ligo and future gravitational wave detectors.
\newblock \emph{Phys. Rev. D}, 101:102005, May 2020.
\newblock \doi{10.1103/PhysRevD.101.102005}.

\bibitem{Cao:20}
H.~T. Cao, et~al.
\newblock High dynamic range thermally actuated bimorph mirror for
  gravitational wave detectors.
\newblock \emph{Appl. Opt.}, 59(9):2784--2790, Mar 2020.
\newblock \doi{10.1364/AO.376764}.

\bibitem{Fricke_2012}
T.~T. Fricke, et~al.
\newblock {DC} readout experiment in {Enhanced} {LIGO}.
\newblock \emph{Classical and Quantum Gravity}, 29(6):065005, feb 2012.
\newblock \doi{10.1088/0264-9381/29/6/065005}.

\end{thebibliography}
